\documentclass[%
 aip,
 amsmath,amssymb,
 reprint,%
]{revtex4-2}

\usepackage{graphicx}% Include figure files
\usepackage{dcolumn}% Align table columns on decimal point
\usepackage{bm}% bold math
\usepackage{xcolor}

\usepackage{booktabs}
\usepackage{makecell}

\usepackage[utf8]{inputenc}
\usepackage[T1]{fontenc}
\usepackage{mathptmx}
\usepackage{textcomp,gensymb} % for \degree

\usepackage{hyperref} % usually, this should be the last package
\hypersetup{
    colorlinks=true,
    linkcolor=blue,
    filecolor=magenta,      
    urlcolor=cyan,
}
\begin{document}

% \preprint{AIP/123-QED}

\title{Detailed characterization  of laboratory magnetized super-critical collisionless shock and of the associated proton energization}
% Force line breaks with \\

\author{W. Yao}
\email{yao.weipeng@polytechnique.edu}
\affiliation{LULI - CNRS, CEA, UPMC Univ Paris 06 : Sorbonne Universit\'e, Ecole Polytechnique, Institut Polytechnique de Paris - F-91128 Palaiseau cedex, France}
\affiliation{Sorbonne Universit\'e, Observatoire de Paris, Universit\'e PSL, CNRS, LERMA, F-75005, Paris, France}

\author{A. Fazzini}
\affiliation{LULI - CNRS, CEA, UPMC Univ Paris 06 : Sorbonne Universit\'e, Ecole Polytechnique, Institut Polytechnique de Paris - F-91128 Palaiseau cedex, France}

\author{S. N. Chen}
\affiliation{ELI-NP, "Horia Hulubei" National Institute for Physics and Nuclear Engineering, 30 Reactorului Street, RO-077125, Bucharest-Magurele, Romania}

\author{K. Burdonov}
\affiliation{LULI - CNRS, CEA, UPMC Univ Paris 06 : Sorbonne Universit\'e, Ecole Polytechnique, Institut Polytechnique de Paris - F-91128 Palaiseau cedex, France}
\affiliation{Sorbonne Universit\'e, Observatoire de Paris, Universit\'e PSL, CNRS, LERMA, F-75005, Paris, France}
\affiliation{IAP, Russian Academy of Sciences, 603155, Nizhny Novgorod, Russia}

\author{P. Antici}
\affiliation{INRS-EMT, 1650 boul, Lionel-Boulet, Varennes, QC, J3X 1S2, Canada}

\author{J. B\'eard}
\affiliation{LNCMI, UPR 3228, CNRS-UGA-UPS-INSA, Toulouse 31400, France}

\author{S. Bola\~{n}os}
\affiliation{LULI - CNRS, CEA, UPMC Univ Paris 06 : Sorbonne Universit\'e, Ecole Polytechnique, Institut Polytechnique de Paris - F-91128 Palaiseau cedex, France}

\author{A. Ciardi}
\affiliation{Sorbonne Universit\'e, Observatoire de Paris, Universit\'e PSL, CNRS, LERMA, F-75005, Paris, France}

\author{R. Diab}
\affiliation{LULI - CNRS, CEA, UPMC Univ Paris 06 : Sorbonne Universit\'e, Ecole Polytechnique, Institut Polytechnique de Paris - F-91128 Palaiseau cedex, France}

\author{E.D. Filippov}
\affiliation{JIHT, Russian Academy of Sciences, 125412, Moscow, Russia}
\affiliation{IAP, Russian Academy of Sciences, 603155, Nizhny Novgorod, Russia}

\author{S. Kisyov}
\affiliation{ELI-NP, "Horia Hulubei" National Institute for Physics and Nuclear Engineering, 30 Reactorului Street, RO-077125, Bucharest-Magurele, Romania}

\author{V. Lelasseux}
\affiliation{LULI - CNRS, CEA, UPMC Univ Paris 06 : Sorbonne Universit\'e, Ecole Polytechnique, Institut Polytechnique de Paris - F-91128 Palaiseau cedex, France}

\author{M. Miceli}
\affiliation{Universit\`a degli Studi di Palermo, Dipartimento di Fisica e Chimica E. Segr\`e, Piazza del Parlamento 1, 90134 Palermo, Italy}
\affiliation{INAF–Osservatorio Astronomico di Palermo, Palermo, Italy}

\author{Q. Moreno}
\affiliation{University of Bordeaux, Centre Lasers Intenses et Applications, CNRS, CEA, UMR 5107, F-33405 Talence, France}
\affiliation{ELI-Beamlines, Institute of Physics, Czech Academy of Sciences, 5 Kvetna 835, 25241 Dolni Brezany, Czech Republic}

\author{V. Nastasa}
\affiliation{ELI-NP, "Horia Hulubei" National Institute for Physics and Nuclear Engineering, 30 Reactorului Street, RO-077125, Bucharest-Magurele, Romania}

\author{S.~Orlando}\affiliation{INAF–Osservatorio Astronomico di Palermo, Palermo, Italy}

\author{S. Pikuz}
\affiliation{JIHT, Russian Academy of Sciences, 125412, Moscow, Russia}
\affiliation{NRNU MEPhI, 115409, Moscow, Russia}

\author{D. C. Popescu}
\affiliation{ELI-NP, "Horia Hulubei" National Institute for Physics and Nuclear Engineering, 30 Reactorului Street, RO-077125, Bucharest-Magurele, Romania}

\author{G. Revet}
\affiliation{LULI - CNRS, CEA, UPMC Univ Paris 06 : Sorbonne Universit\'e, Ecole Polytechnique, Institut Polytechnique de Paris - F-91128 Palaiseau cedex, France}

\author{X. Ribeyre}
\affiliation{University of Bordeaux, Centre Lasers Intenses et Applications, CNRS, CEA, UMR 5107, F-33405 Talence, France}

\author{E. d'Humi\`eres}
% \email{emmanuel.dhumieres@u-bordeaux.fr}
\affiliation{University of Bordeaux, Centre Lasers Intenses et Applications, CNRS, CEA, UMR 5107, F-33405 Talence, France}

\author{J. Fuchs}
% \thanks{julien.fuchs@polytechnique.edu}
% \email{julien.fuchs@polytechnique.fr}
\affiliation{LULI - CNRS, CEA, UPMC Univ Paris 06 : Sorbonne Universit\'e, Ecole Polytechnique, Institut Polytechnique de Paris - F-91128 Palaiseau cedex, France}

\date{\today}% It is always \today, today,

\begin{abstract}

Collisionless shocks are ubiquitous in the Universe and are held responsible for the production of non-thermal particles and high-energy radiation.
In the absence of particle collisions in the system, theoretical works show that the interaction of an expanding plasma with a pre-existing electromagnetic structure (as in our case) is able to induce energy dissipation and allow for shock formation. Shock formation can alternatively take place when two plasmas interact, through
microscopic instabilities inducing electromagnetic fields which are able in turn to mediate energy dissipation and shock formation. 
Using our platform where we couple a fast expanding plasma induced by high-power lasers (JLF/Titan at LLNL and LULI2000) with high-strength magnetic fields, we have investigated the generation of magnetized collisionless shock and the associated particle energization. 
We have characterized the shock to be collisionless and 
% super-Alfv\'{e}nic
super-critical.
We report here on measurements of the plasma density, temperature, the electromagnetic field structures, and particle energization in the experiments, under various conditions of ambient plasma and B-field. 
We have also modelled the formation of the shocks using macroscopic hydrodynamic simulations and the associated particle acceleration using kinetic particle-in-cell simulations. 
As a companion paper of \citet{yao2020laboratory}, here we show additional results of the experiments and simulations, providing more information to reproduce them and demonstrating the robustness of our interpreted proton energization mechanism to be shock surfing acceleration.

\end{abstract}

\maketitle

\section{\label{sec:intro}Introduction}

The acceleration of energetic charged particles by collisionless magnetized shock is a ubiquitous phenomenon in astrophysical environments, among which the most energetic particles are the ultra-high-energy cosmic rays (UHECRs) accelerated in the interstellar medium (ISM) \cite{2009Sci...325..719H,2013Sci...340...45N}.
In this case, the source of collisionless dissipation is self-generated electromagnetic fields, resulting from kinetic instabilities such as the Weibel one.
Besides, particles are also accelerated in our solar system due to collisionless magnetized shocks developed by the interaction of the solar wind with planetary magnetospheres \cite{turner,amano2020observational} and, at larger distances, with the ISM \cite{decker2008voyager2}.
In that case, the source of collisionless dissipation is the pre-existing global electromagnetic structure. This will be the case for the experiment detailed here, where we apply a global strong magnetic field onto a laser-ablated fast plasma. 
Since these shocks usually have their Magnetosonic Mach number $M_{ms} = v_{sh} / v_{ms} \gtrsim 2.7$  (where $v_{sh}$ is the shock velocity, $v_{ms} = \sqrt{C_s^2 + v_A^2}$ is the Magnetosonic velocity, $C_s$ and $v_A$ are the ion sound velocity and Alfv\'{e}nic velocity, respectively), they belong to the so-called super-critical regime \cite{coroniti1970dissipation,edmiston1984parametric}, which means the shock is not maintained by classical dissipation means alone. In order to help maintain a shock, the additional channel to expel energy is achieved by reflecting particles back upstream \cite{balogh2013physics}.

A variety of acceleration mechanisms have been evoked as a way to transfer the energy from the shock waves to the particles, including shock surfing acceleration (SSA), shock drift acceleration (SDA), and diffusive shock acceleration (DSA). DSA requires high initial energy before further acceleration \cite{zank1996interstellar}, thus raising the so-called ``injection problem'' \cite{lembege2004selected}; while SSA and SDA are believed to be responsible for generating the pre-accelerated seed particles, i.e. for the initial accelerating process from thermal energies. 
Although it is still under debate whether SSA or SDA dominates the pre-acceleration process in various collisionless shock environments \cite{burrows2010pickup,zank2009microstructure,chalov2016acceleration}, we can distinguish them by the following two aspects: On the one hand, in SSA, charged particles first get reflected at the shock front (due to the cross-shock potential electric field), then they surf along the shock front against the convective electric field ($\bm{E} = - \bm{v} \times \bm{B}$), and thus they gain energy. While in SDA, charged particles drift (due to the magnetic field gradient at the shock front) along the convective electric field and then gain energy \cite{guo2013acceleration}. On the other hand, SSA requires a thin shock width, compared to the Larmor radius of the charged particles, while SDA needs the opposite (so that the charged particles can gyrate and drift within the shock layer) \cite{zank1996interstellar,yang2009shock}.

However, because of the immense spatial scales involved with collisionless phenomenon (e.g. the mean-free-path is $\lambda_{mfp} \sim 1$ AU in the Solar system), only a very small sampling of the shock formation and dissipation mechanisms can be realized. As a result, we still do not have a full understanding of the formation and evolution of collisionless shocks, and the question of the effectiveness and relative importance of SDA and SSA is still largely debated in the literature \cite{yang2012impact}.
To further our understanding, laboratory experiments (and their simulations) have been proven to be an effective tool, providing highly-resolved, reproducible and controllable multi-dimensional datasets that can complement astrophysical observations \cite{paul2006high,lebedev2019exploring}. 
Below, we will now briefly review the investigation of collisionless shocks via laboratory experiments. 

The route that has been up to now most explored in the laboratory is to produce a shock (mediated by the Weibel filamentation instability) by colliding two ablative, unmagnetized flows driven by high-energy nanosecond lasers. This setup has yielded promising results at the Omega Laser Facility \cite{fox2013filamentation,huntington2015observation,park2015collisionless} and the National Ignition Facility (NIF) \cite{park2016laboratory,ross2017transition}, as well as at many other laser facilities all over the world \cite{courtois2004experiment,yuan2018laboratory, kuramitsu2011time}. Recently, experiments on collisionless shocks in plasma flows in which there was significant self-generated magnetic field showed, for the first time, the formation of magnetized collisionless shock, with the generation of Weibel instability and observation of electron acceleration in the turbulent structure \cite{li2019collisionless}. 
Most recently, the dynamics of the ion Weibel instability has been characterized by local, quantitative measurements of ion current filamentation and magnetic field amplification in interpenetrating plasmas via optical Thomson scattering (TS) \cite{swadling2020measurement}. What's more, the generation of sub-relativistic shocks, together with relativistic electron acceleration, has been demonstrated to be within the reach of larger-scale, NIF-class laser systems \cite{fiuza_electron_2020}. 

Another setup relies on a plasma expanding into a pre-formed ambient magnetized secondary plasma. Thanks to the magnetisation, the target ions create a collisionless magnetic piston that accelerates the ambient plasma to super-Alfv\'{e}nic velocity, thus creating a high-Mach number shock with velocity of the order of 1000 km/s \cite{schaeffer2012generation,niemann2014observation,schaeffer2017generation,schaeffer2017high}. Recently, Schaeffer et al. have been able to make significant progress in characterizing the formation of collisionless shocks in terms of ion and electron density and temperature, as well as electric and magnetic field strengths as a function of time at OMEGA \cite{schaeffer2019direct}. 

Besides, at the LULI laser facility at \'Ecole Polytechnique (France), collisionless shock waves and ion-acoustic solitons have been investigated by proton radiography \cite{romagnani2008observation}. Moreover, significant electron pre-heating via lower-hybrid waves was also achieved in laboratory laser-produced shock experiments with strong magnetic field, providing a potential mechanism for the famous ``injection'' problem \cite{rigby2018electron}. Additionally, at the VULCAN laser facility at the Rutherford Appleton Laboratory, the temporally and spatially resolved detection of the forming of a collisionless shock was achieved \cite{ahmed2013time}.

In contrast to the above schemes, novel setups have been used with ultra-high-intensity lasers. For example, at the XingGuang III laser facility at the Laser Fusion Research Center in China, using a short (2 ps) intense (10$^{17}$ W/cm$^2$) laser pulse, an electrostatic (ES) collisionless shock, together with the filaments induced by ion-ion acoustic instability, could be observed via proton radiography \cite{jiao2019experimental}.

In our experimental campaigns at JLF/Titan and LULI2000 \cite{yao2020laboratory}, we investigated shock formation combining laser-produced plasmas, a background medium and a strong ambient magnetic field (as detailed below). We chose to have an expanding plasma to drive a shock into an ambient gas in the presence of a strong external magnetic field. Contrary to \citet{schaeffer2017generation}, in our setup, the expanding plasma and the magnetic field were decoupled as the higher Z piston evacuates the magnetic field and was thus unmagnetized. This also allowed us to  simultaneously have a highly magnetized ambient plasma (with homogeneous and steady magnetic field) and a high-$\beta$ piston ($\beta \equiv P_{thermal} / P_{mag}$ is the ratio of the thermal pressure to the magnetic one). 
% \yao{(2) to be able to vary the B-field independently of the expanding plasma. This later point allowed us, in particular, to show the essential role played by the magnetic field in allowing steep shock generation}. 
Moreover, since our magnetic field strength was more than two times higher \cite{albertazzi2013production}, reaching 20 T comparing to the 8 T in \citet{schaeffer2017generation}, we were able to decouple more strongly the electrons from the ions \cite{yao2019kinetic}, and the shock was able to fully separate from the piston, which is crucial for its characterization \cite{schaeffer2020kinetic}. As a result, we have been able to characterize the plasma density, temperature, as well as the E-field developed at the shock front, and more importantly, observe strong non-thermal accelerated proton populations for the first time. 

In this paper, we will first show that laboratory experiments can be performed to generate and characterize globally mildly super-critical, quasi-perpendicular magnetized collisionless shocks in Section~\ref{experiments}, and detail their characteristics. 
Then, we will detail  in Section~\ref{3D-MHD} three-dimensional (3D) magneto-hydrodynamic (MHD) simulations reproducing the laser-driven piston generation and the following shock formation process.
In Section~\ref{1D-PIC}, with the parameters characterized in the experiment, we will report the results of kinetic particle-in-cell (PIC) simulations, which pinpoint that shock surfing acceleration (SSA) can be effective in energizing protons from the background plasma to hundred keV-level energies. 

\section{Experimental setup and results}\label{experiments}

\subsection{Experimental setup}

The experiments were performed at the JLF/Titan (LLNL, USA) and LULI2000 (France) laser facilities with similar laser conditions but using complementary diagnostics, which was mostly linked with the availability of different auxiliary laser beams at each facility. 

\begin{figure}[htp]
    \centering
   \includegraphics[width=0.45\textwidth]{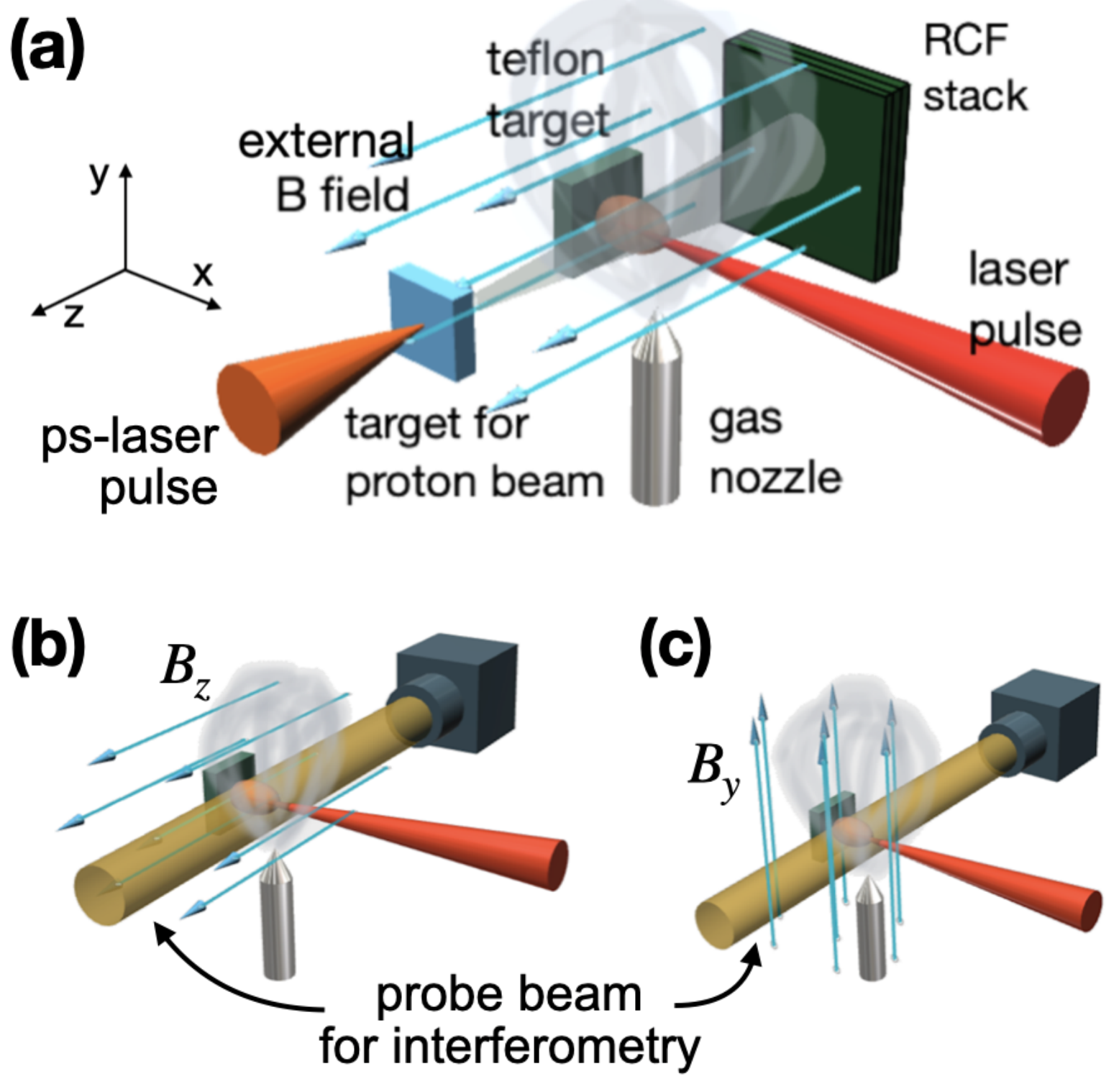}
    \caption{\textbf{Experimental setup and diagnostics used to characterize a magnetized shock.} Proton radiography and interferometry diagnostics have been used alternatively along the axis perpendicular to the laser and to the plasma flow (i.e. the z-axis). (a) Proton radiography setup. (b-c) In the case of interferometry, we could rotate the coil in order to have two different magnetic field orientations with respect to the field of view of the probe beam.}
    \label{fig:exp_setup}
\end{figure}

In the experiment at JLF/Titan, as is shown in Fig.~\ref{fig:exp_setup}, the collisionless shock was generated by sending a plasma, generated by having a high-power laser (1 $\mu$m wavelength, 1 ns duration, 70 J energy,  and $1.6 \times 10^{13}$ W/cm$^2$ on-target intensity) irradiating a solid target (Teflon, CF$_2$), into a low-density ($10^{18}$ cm$^{-3}$) H$_2$ ambient gas pulsed from a nozzle prior to the shot, and in the presence of a 20 T magnetic field that is homogeneous and steady-state at the time scale of the experiment. As shown in Fig.~\ref{fig:exp_setup}, the magnetic field created by a Helmholtz coil system \cite{higginson2019laboratory, albertazzi2013production} is oriented along the y- or z-axis.

\subsection{Density characterization through optical interferometry}

\begin{figure}[htp]
    \centering    
    \includegraphics[width=0.45\textwidth]{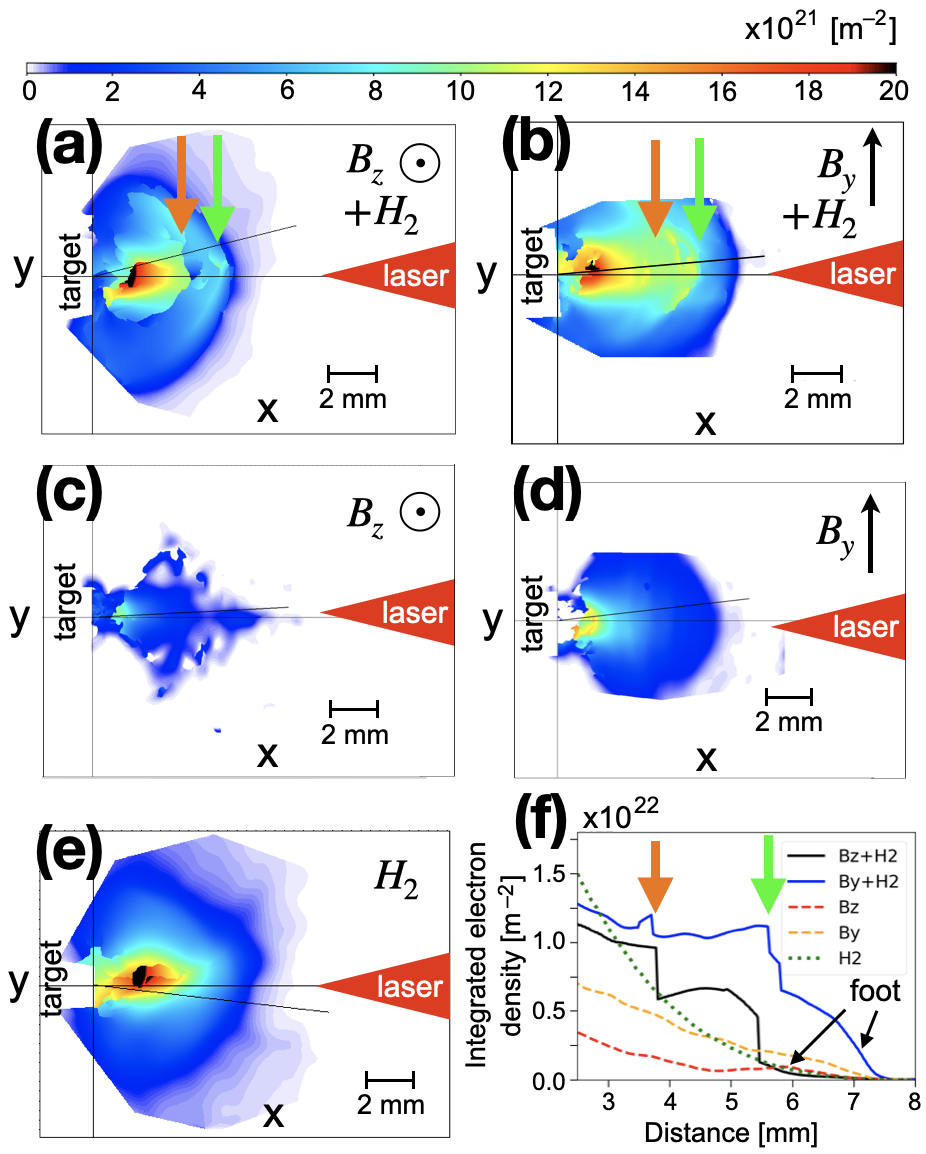}
    \caption{\textbf{Integrated plasma electron density, as measured by optical probing at 4 ns after the laser irradiation of the target, in three different cases.} 
    (a) and (b) Cases with both ambient gas and B-field in the xy- and xz-plane, respectively.  
    (c) and (d) Cases with only B-field but without ambient gas \cite{khiar2019laser,filippov2020enhanced} in the xy- and xz-plane, respectively.
    (e) Cases with only ambient gas but without B-field in the xy-plane (the xz-plane will be the same). 
    Each image corresponds to a different laser shot, while the color scale shown at the top applies to all images.
    (f) The lineouts along the thin dark lines shown in each image. 
    The laser comes from the right side and the piston source target is located at the left (at $x=0$). Yellow arrows indicate the piston front, while green arrows indicate the shock front.
    }
    \label{fig:optical_probe}
\end{figure}

Using an interferometry setup \cite{higginson2017detailed}, the plasma electron density is recorded by optically probing the plasma (with a mJ, 1~ps auxiliary laser pulse). In Fig.~\ref{fig:optical_probe}, we present the overall electron density recorded in three different cases.

For the case with both ambient gas and B-field shown in Fig.~\ref{fig:optical_probe} (a) and (b), the laser irradiation induced the expansion of a hot plasma (the piston) that propagates along the x-axis and the collisionless shock is formed as a consequence of the plasma piston propagating in the magnetized ambient gas \cite{schaeffer2017generation}. We can clearly see both the piston front and the shock front (indicated by the orange and green arrows, respectively), and indeed they are well detached from each other, enabling us to characterize them separately.

A lineout of the plasma density is shown in Fig.~\ref{fig:optical_probe} (f), where the piston and shock fronts are also well identified by the abrupt density changes. The piston front is steepened by the compression of the magnetic field (see also below). Besides, we can clearly see a ``foot'' structure ahead of the shock front in the upstream (US) region for the case with both ambient gas and B-field, indicating the formation of the magnetized shock \cite{giagkiozis2017validation}.

In contrast, for the case with only B-field but without ambient gas \cite{khiar2019laser} shown in Fig.~\ref{fig:optical_probe} (c) and (d), 
% the piston behaves in an asymmetric way in xy- and xz-plane, similar to that in \citet{khiar2019laser}. 
due to the lack of ambient gas, no collisionless shock is formed ahead of the piston.
For the case with only ambient gas but without B-field in Fig.~\ref{fig:optical_probe} (e), no shock is formed as well in the ambient gas. From the corresponding lineout in (f), it is clear that only a smooth plasma expansion into the ambient (the green dashed line) can then be seen.

\subsection{Piston compression characterization through X-ray spectroscopy}

To further characterize the piston, the x-ray ion emission of Fluorine compressed within the expanding piston was measured by a Focusing Spectrometer with high Spatial Resolution (FSSR) \cite{Faenov:1994} at both laser facilities. It was based on a spherically-bent mica ($2d = 19.9376$~\r{A}) crystal with a curvature radius of $R = 150$~mm. Spatial resolution of 100 $\mu$m per pixel was achieved along the plasma expansion. Image Plate (Fujifilm TR BAS) was used as a fluorescent detector. The implemented scheme resulted in 13-16~\r{A} spectral range with a high resolution ($\lambda$/d$\lambda$ is higher than 1000). It covers spectral lines of Fluorine: resonance H-like (2p--1s transition) and He-like (3p--1s, 4p--1s, 5p--1s etc.) transitions as well as dielectronic satellites to Ly$_{\rm \alpha}$. The diagnostic allowed us to measure electron density and temperature profiles of the piston expansion using a quasi-stationary approach \cite{Ryazantsev2015}. The method is based on analysing the relative intensities of spectral lines of the same charge state and also takes into account the recombining plasma with a “frozen” ion charge.
\begin{figure}[htp]
    \centering
    \includegraphics[width=0.4\textwidth]{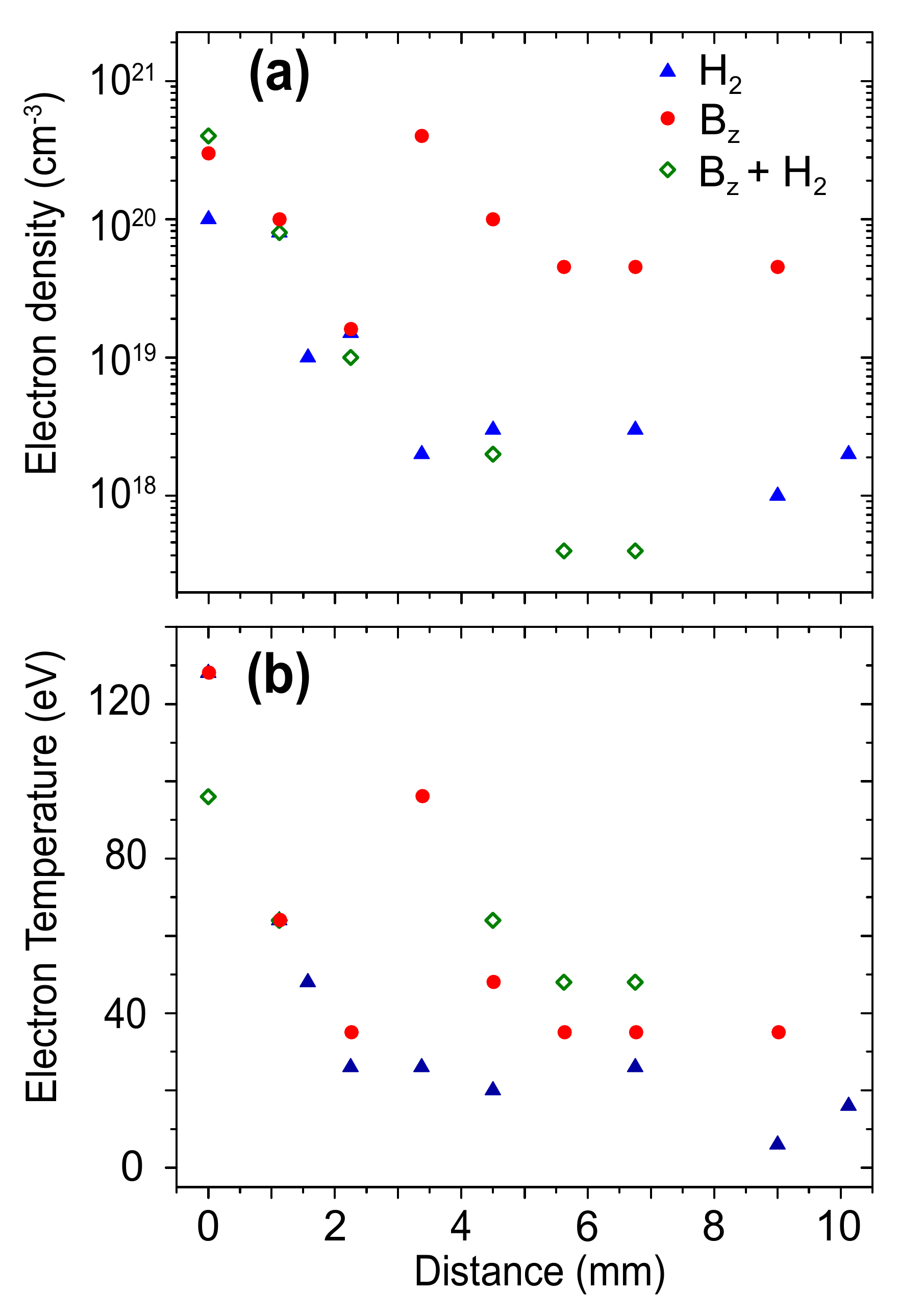}
    \caption{\textbf{FSSR evaluation of (a) electron density and (b) electron temperature of the laser-produced piston in three different configurations (see legend) along the expansion axis.} The measurements are based on the analysis \cite{Filippov2019} of the relative intensities of the x-ray emission lines of He-like and H-like (see text) Fluorine ions in the expanding plasma in the range of 13-16~\r{A}. The quasi-stationary \cite{Ryazantsev2015} approach was applied for He-like series of spectral lines assuming a "frozen" ion charge state. The 0 point corresponds to the target surface. The spatial resolution of about 100 $\mu$m was achieved. The signal is time-integrated.
    }
    \label{fig:FSSR}
\end{figure} 

Figure~\ref{fig:FSSR} (a) shows that obviously the piston encounters stronger hindrance in the case with both ambient gas (H$_2$) and B-field ($B_z$) (see the green diamonds), comparing with other cases (i.e. the case with only $B_z$ in red dots and the case with only H$_2$ in blue triangles).
We also see in Fig.~\ref{fig:FSSR} (b) that the electron temperature in the case of $B_z + H_2$ becomes the highest at the piston front (between 4 and 7 mm), comparing with other cases.
In addition, at the position of 4.5 mm, the evaluated electron density for the case of $B_z + H_2$ is around $2-3 \times 10^{18}$ cm$^{-3}$ and the electron temperature is about 65 eV, which are well-reproduced by our FLASH simulations, see Fig.~\ref{fig:FLASH_2ns_Ne_Te_Ti} (a) and (b). 

\subsection{Electric field characterization through proton radiography}

The single shock front was also probed with protons in order to measure the local electric field. The probing protons (accelerated by the Target Normal Sheath Acceleration process \cite{wilks1992absorption} from an auxiliary target and using the short-pulse arm of Titan) was sent parallel to the B-field, i.e. along the z-axis, as is shown in Fig.~\ref{fig:exp_setup} (a). 
% Probing along another direction would be impossible, due to the bending it would impair on the protons.

\begin{figure}[htp]
    \centering
    \includegraphics[width=0.45\textwidth]{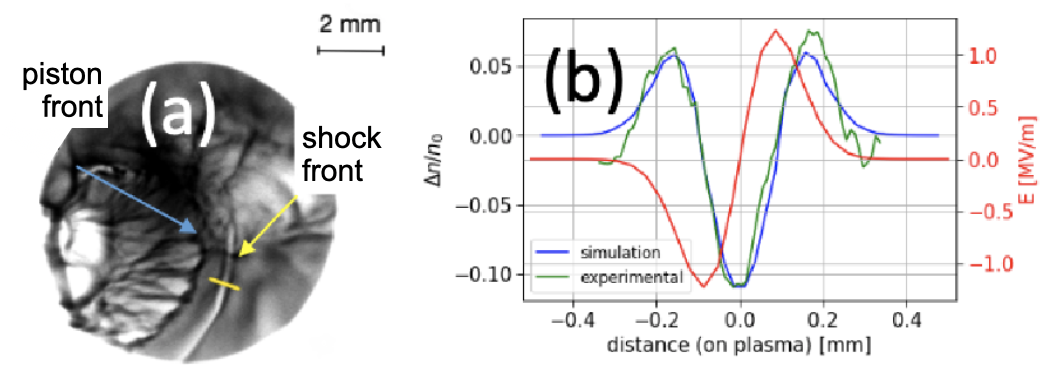}
    \caption{\textbf{Proton radiography} (collected on a RCF film and employing 19 MeV protons) of the same configuration as shown in Fig.~\ref{fig:optical_probe} (a), 5 ns after the laser pulse. (b) Lineout of the proton modulation along the yellow line indicated in (a). The proton modulation recorded at the shock front can be fitted by a bipolar electric field, as shown in the red solid line.
    }
    \label{fig:ProtonRadio}
\end{figure}

As shown in Fig.~\ref{fig:ProtonRadio} (a), we could clearly observe the same structures of the piston front and the shock front, consistent with those observed via optical probe, as shown in Fig.~\ref{fig:optical_probe}. 
By analysing the proton deflection structure, we could infer that we had a bipolar electric field at the shock front, with a total width of 0.4 mm and an amplitude of around 1 MV/m, as shown in Fig.~\ref{fig:ProtonRadio} (b). To get this result, we imposed a certain 3D electric field map and simulated the proton dose that we would get on a detector. The electric field had been modulated in order to obtain a simulated dose (blue line) matching as much as possible the measured one (green line). Note that the proton deflection structure is accumulated along the z-direction. We will compare it with the particle-in-cell simulation results and discuss it in detail in Sec.~\ref{1D-PIC}.
%From it, as shown in Fig.~\ref{fig:ProtonRadio} (b), we can infer that the E-field at the shock front had a double jump structure, with a field reversal. 

\begin{figure}[htp]
    \centering
    \includegraphics[width=0.4\textwidth]{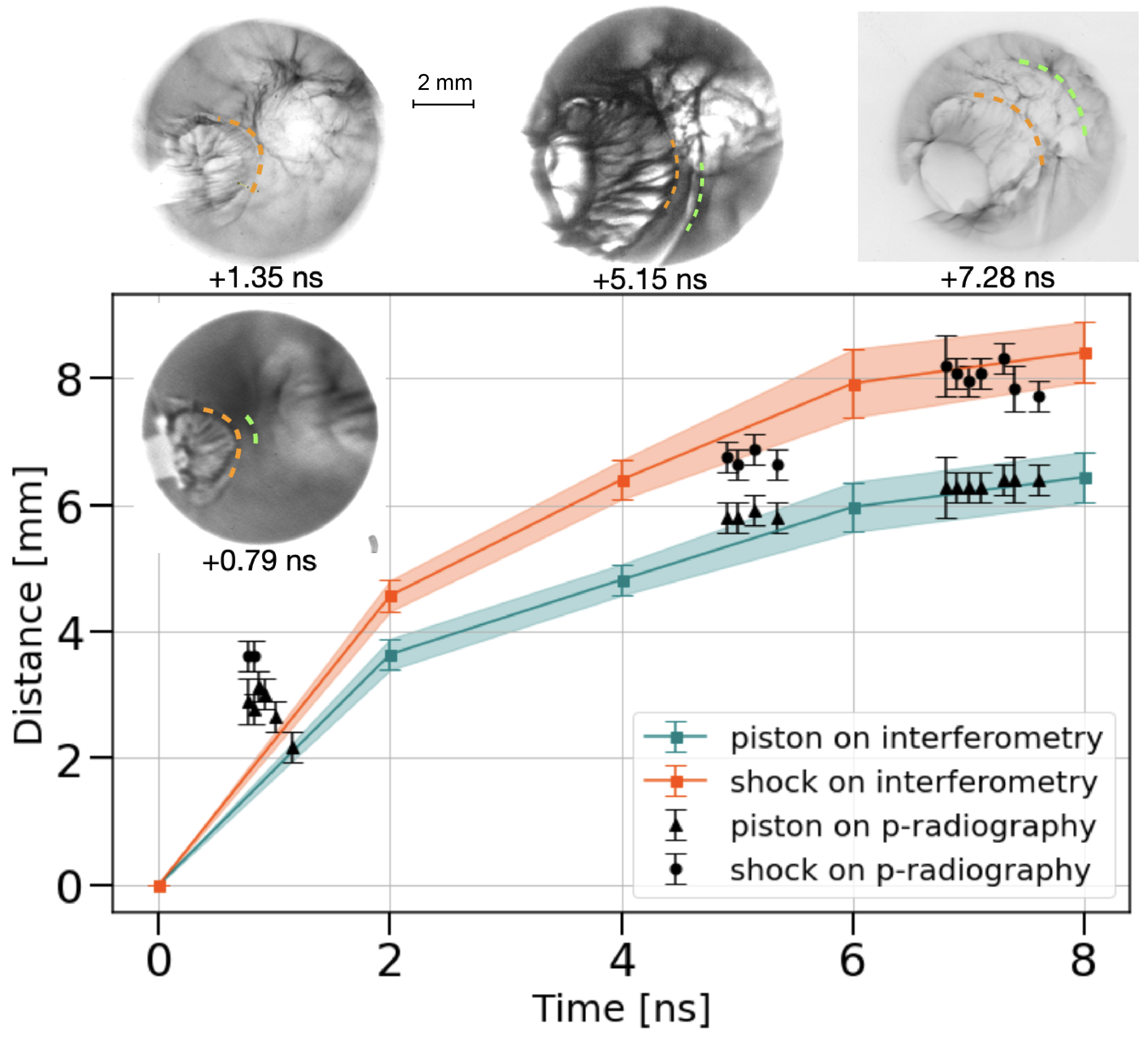}
    \caption{\textbf{Piston and shock front position over time on the electron density (via interferometry) and on the electric field (via proton radiography).}  Images of proton radiography doses at different times are also shown, with dashed lines for piston front (in orange) and shock front (in green).
    }
    \label{fig:PR_interf}
\end{figure}

Moreover, we compared the position of the shock structures seen in the electron density (via interferometry) with that in the electric field (via proton radiography) for the case with both external B-field and ambient gas. For the former, we have considered the point where the electron density had a sharp jump, as shown in Fig. \ref{fig:optical_probe}f; as for the latter, we have taken into account the external edges of the proton dose accumulation. As is shown in Fig.~\ref{fig:PR_interf}, the evolution of the piston front and the shock front through both diagnostics are illustrated together (see legends for details), and it is clear that the results are quite consistent with each other. Note that when the target was not clearly visible in the radiography, i.e. for the series of points around 5 ns, we made use of the interferometry results to shift all the points of the right amount, while the distances between the piston and the shock fronts were kept constant. The original RCFs for the data points at various times are also shown.

%%%%% TS section start
\subsection{Temperature characterization through Thomson scattering}

\begin{figure}[htp]
    \centering
    \includegraphics[width=0.43\textwidth]{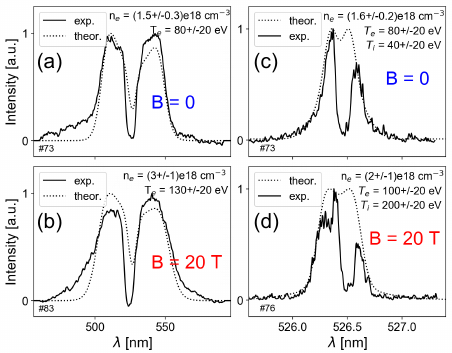}
    \caption{\textbf{Thomson scattering measurements, of the plasma density and temperatures, in the region  downstream of the shock front, and for different cases.} (a) measurement on the electron waves for $B=0$ case (i.e. with only ambient gas), allowing to retrieve the local electron number density and electron temperature, as stated; (b) the same measurement for $B=20$ T case (i.e. with external B-field and ambient gas). (c) measurement on the ion waves in the plasma for $B=0$ case, allowing to retrieve the local electron and ion temperatures, as stated; (d) the same measurement for $B=20$ T case. Solid lines are for experimental data profiles, while dashed lines are for theoretical spectra. The stated uncertainties in the retrieved plasma parameters represent the possible variation of the parameters of the theoretical fit, as well as the shot-to-shot variations observed in the same conditions. 
    }
    \label{fig:TS_diag}
\end{figure}

With a second high-energy auxiliary (0.5 $\mu$m wavelength, 1 ns, 15 J, focused over $\sim$ 40 $\mu$m along the z-axis and propagated throughout the plasma) available at LULI2000, we are able to characterize the plasma temperature by performing Thomson scattering (TS) off the electron and ion waves in the plasma (used in a collective mode \cite{froula2011plasma} and analyzed by different spectrometers). 

Figure~\ref{fig:TS_diag} shows the TS measurements in the region downstream (DS) compared to the shock front for cases with and without the external B-field. By comparing the experimental data profiles with the theoretical equation of the scattered spectrum for coherent TS in unmagnetized and non-collisional plasmas, with the instrumental function taken into an account, we are able to retrieve the local electron number density, as well as the electron and ion temperatures \cite{froula2007quenching}. For the case without the B-field (i.e. with only ambient gas), both TSe and TSi give $n_e \sim 1.5\times 10^{18}$ cm$^{-3}$ and $T_e \sim 80$ eV, and TSi also gives $T_i \sim 40$ eV in the DS region, as can be seen in Fig.~\ref{fig:TS_diag} (a) and (c). However, for the case with $B=20$ T, we see strong compression and heating in the DS region, indicated by the higher density and temperatures, i.e. $n_e \sim 2.5 \times 10^{18}$ cm$^{-3}$, $T_e \sim 115$ eV, and $T_i \sim 200$ eV, as can be seen in Fig.~\ref{fig:TS_diag} (b) and (d). With the characteristic feature of $T_i > T_e$, the effective formation of a shock can be inferred.

%%%%% TS section end

\subsection{Evidence for proton energization}

\begin{figure}[htp]
    \centering
    \includegraphics[width=0.4\textwidth]{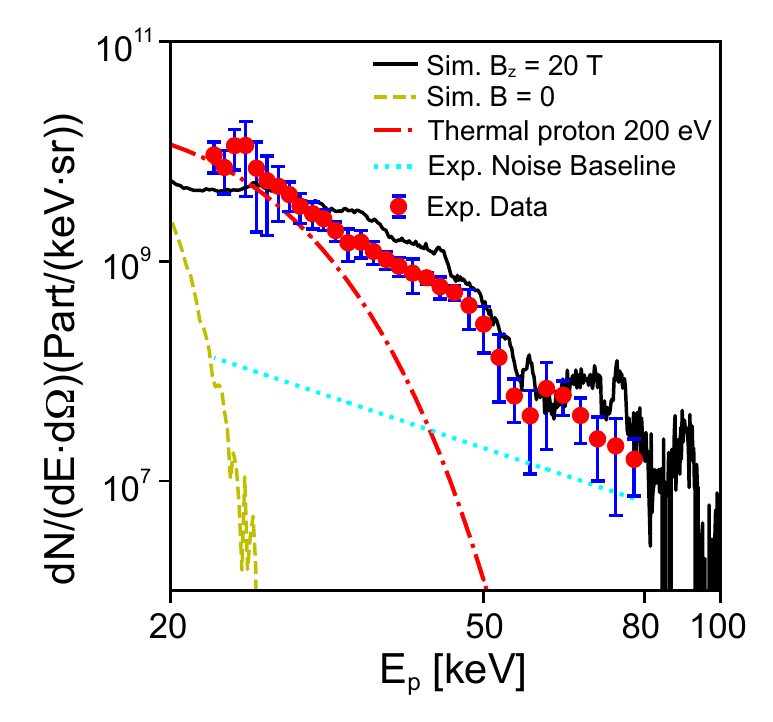}
    \caption{\textbf{Proton energy spectrum.} The experimental data is shown with red dots and blue error bars; the simulation results are shown with black solid line for the case with $B=20$T and yellow dashed line for $B=0$ case; the analytical thermal proton spectrum is shown with red dash-dot line (200 eV); and the experimental noise baseline is shown in cyan dotted line.
}
    \label{fig:ion_spectrometer}
\end{figure}
For the observation of the non-thermal proton spectrum, we use a standard magnetic spectrometer, with permanent magnets of 0.5 T strength. It was located close to the target (17.5 cm away) in order to maximize its collection efficiency, and it had its main axis along z, the main of the external B-field.
% (in an alternate mode to performing TS). 
That spectrometer has been calibrated precisely with a Hall probe and on many previous campaigns using filters to verify its energy dispersion. The protons are deflected by the magnetic field inside the spectrometer and landed after a short drift space onto Imaging plates (of TR type), the detector used here. These detectors are absolutely calibrated \cite{manvcic2008absolute}.

The recorded proton spectrum is shown in Fig.~\ref{fig:ion_spectrometer} with red dots and blue error bars. Comparing it to the analytical thermal proton spectra (200 eV in red dash-dotted lines, as is observed in \cite{yao2020laboratory} through TS), it is clear that the proton energization is non-thermal.
The cutoff energy reaches to about $80$ keV.
% close to the Hillas limit ($E_{HL} < e v R B \sim 100$ keV) \cite{hillas1984origin, drury2012origin} (with the elementary charge $e=1.6\times 10^{-19}$ C , the shock velocity $v = 1.5 \times 10^6$ m/s , the ``acceleration length'' $R \sim 3$ mm, and the B-field strength $B = 20$ T).
% We should note that the Hillas limit is in general an optimistic and global estimation of the maximum energy, based on equating the Larmor radius of the particle to the length of the acceleration region and considering that the acceleration takes place over the entire region. Thus, as an upper limit, it does not describe the underlying acceleration mechanism, which needs numerical simulations to pinpoint (as will be shown below).
Also note that there is no signal recorded above the noise baseline for the case with only the B-field or the ambient gas.

\section{MHD simulations with FLASH}\label{3D-MHD}

We use the 3D MHD code FLASH~\cite{fryxell2000flash} to study the dynamics of the plasma plume expansion and shock formation in the ambient gas with the strong magnetic field, using the same parameters as the JLF/Titan experiment. 
The simulations are initialized in 3D geometry, using three temperatures (two for the plasma, and one for the radiation) with the equation-of-state of \citet{kemp1998equation} and radiative transport, in the frame of ideal MHD and including the Biermann battery mechanism of magnetic field self-generation in plasmas \cite{haines1986magnetic}. 
Specifically, the laser beam is normal to a Teflon target foil and has an on-target intensity of $10^{13}$ W/cm$^2$; the generated plasma plume expands in the hydrogen gas-jet having an uniform density of $10^{18}$ cm$^{-3}$. Moreover, the plasma plume expands in the uniform external magnetic field of 20 Tesla (aligned along the z-axis, as in the experiment).

\begin{figure*}[htp]
    \centering
    \includegraphics[width=0.8\textwidth]{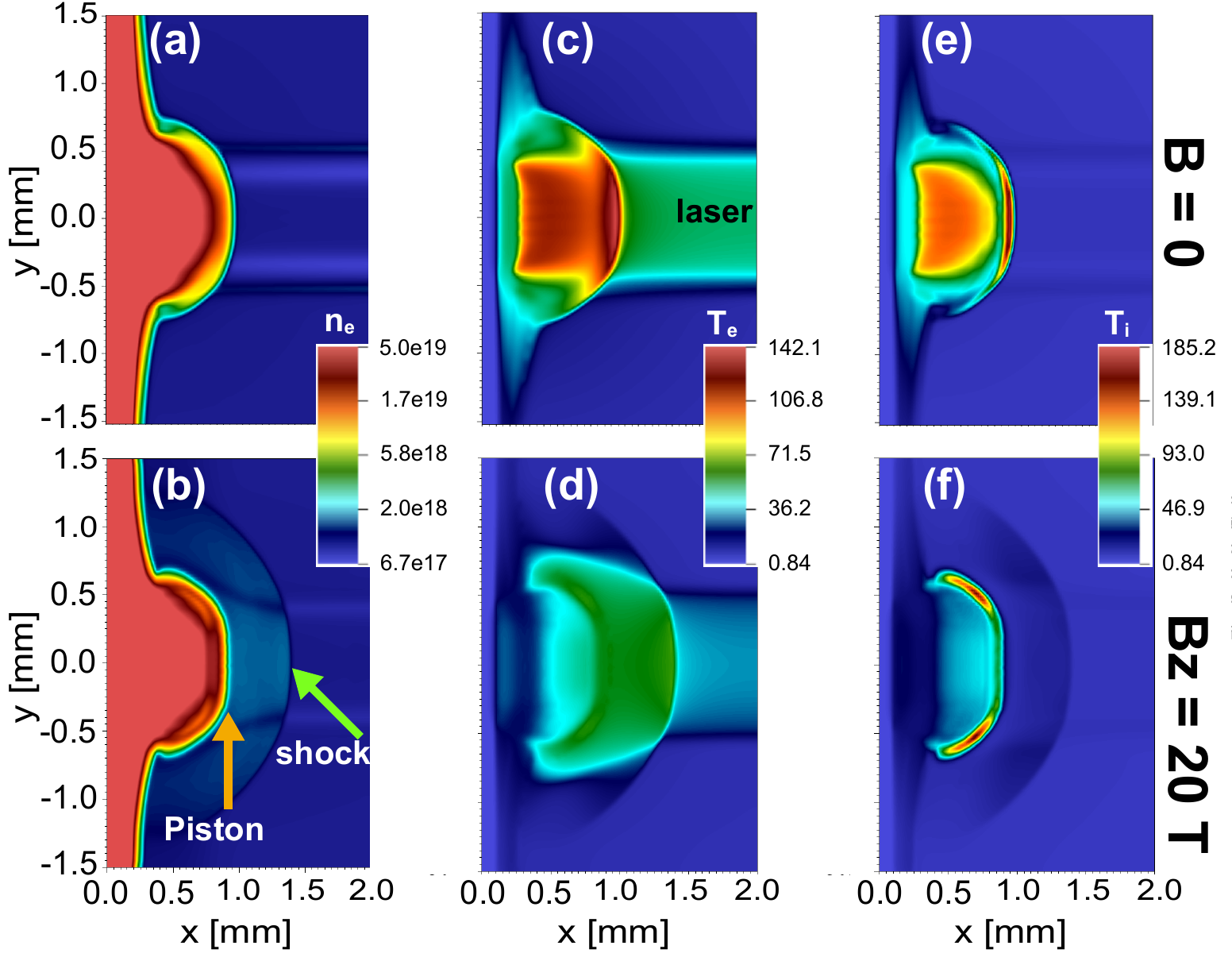}
    \caption{\textbf{FLASH simulation investigating a single shock formation and performed in the conditions of the JLF/Titan experiment.} Maps extracted from FLASH simulations at 2 ns (after the laser irradiation) of: (a) and (b) electron density, $n_e$ in cm$^{-3}$, (c) and (d) electron temperature, $T_e$ in eV, (e) and (f) ion temperature, $T_i$ in eV. The upper row is for the case without B-field, while the lower row is for the case with B-field. All maps are in linear scale. This XY-plane slice is cut at Z=0. The laser comes from the right side along $y=0$, and the target is at the left side. The yellow arrow indicates the piston edge, while the green arrow indicates the shock front. As FLASH cannot tolerate vacuum, we do not have the FLASH simulation for the case with only B-field but without ambient gas. 
    }
    \label{fig:FLASH_2ns_Ne_Te_Ti}
\end{figure*}

Figure~\ref{fig:FLASH_2ns_Ne_Te_Ti} shows the FLASH simulation results, i.e. the electron density, electron temperature and ion temperature from FLASH at t = 2 ns (after the laser irradiation), in two different cases (the upper row is for the case with only ambient gas but without B-field, while the lower row is for the case with both the ambient gas and the B-field). 
We can observe that the structures of both the hydrodynamic piston and the induced shock, which propagates inside the ambient, are qualitatively reproduced compared to the experiment. 
The Teflon expanding piston produces a forward shock in the ambient (around $x=1.4$ mm), as well as a reverse shock inside the Teflon piston (around $x=0.8$ mm). 
The electron density is $\sim1.6\times10^{18}$ cm$^{-3}$ in the forward shock in the gas and increases up to $\sim~5\times10^{19}$ cm$^{-3}$ in the reverse shock. 
The electron temperatures are between 60 to 70 eV in the forward and reverse shocks. 
Both correspond quite well to what is measured in the experiment (see the FSSR measurements in Fig.~\ref{fig:FSSR} and the TS measurements in \cite{yao2020laboratory}).
The ion temperature is 15 eV in the forward shock and between 80 eV and 180 eV inside the reverse shock. 

Concerning the electron temperature, the FLASH simulation results are two times lower compared to the TS measurements in the DS region shown in Fig.~\ref{fig:TS_diag}; while for the ion temperature, the situation is worse as it is ten times less in the forward shock compared to the TS measurements. Also note that we have not seen the foot structure ahead of the shock in the FLASH simulations.
Such discrepancies between the MHD simulations and the experiments show the difficulties to reproduce the shock condition in our case. This points to the fact that the shock evolution is dominated by kinetic effects. This is why we have resorted to using PIC simulations, the initial conditions of which are taken from the experimental measurements. Nevertheless, we can still observe that the FLASH simulations reproduce well the dynamics of the piston that induces the shock.

Since FLASH has the ability to model magnetic field generation through the Biermann battery effect, it allows us to assess the importance of this effect in the present configuration. Biermann battery generation of magnetic field is typically important only close to the target surface (order of 1 mm), and it is localized over the steep temperature gradients generated by the laser beam and rapidly decays once the laser beam is off (see for example \cite{li_2006,Gao2015,cecchetti_2009}). As the shock is induced by the piston in the ambient gas 1 mm away from the target surface after the laser is off ($\sim 2$ ns), as shown in Fig.~\ref{fig:FLASH_2ns_Ne_Te_Ti}, the Biermann battery effect is negligible, compared to that of the strong externally applied B-field.

\section{Kinetic simulations with SMILEI}\label{1D-PIC}

The proton energization via the collisionless shock is modelled with the kinetic PIC code SMILEI \cite{derouillat2018smilei}. 
During the interaction between the shock front and the ambient plasma, as the scale across the shock ($\sim$mm) is much larger than that along the shock ($\sim \mu$m), we can treat this quasi one-dimensional (1D) interaction via the 1D3V version of the code.

\begin{figure}[htp]
    % \centering
    \includegraphics[width=0.45\textwidth]{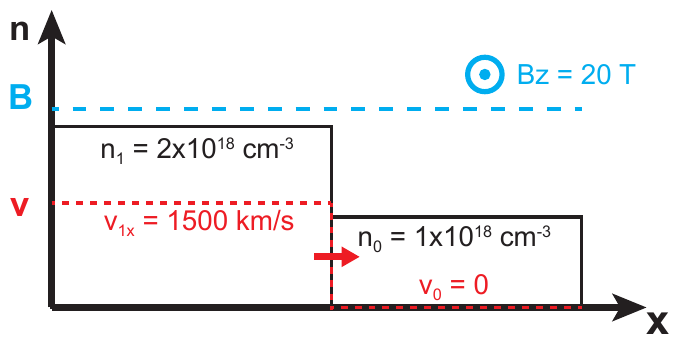}
    \caption{\textbf{Diagram of PIC simulation setup.} The shocked plasma lies on the left half of the simulation box, drifting towards right; while the ambient plasma lies on the right half. Number density (n, black solid), drifting velocity (v, red dotted), and the magnetic field (B, blue dashed) are noted with their value. We stress here that the shock width is initialized to be equal to the ion inertial length $d_i = 200\ \mu$m.}
    \label{fig:simu_setup}
\end{figure}

As is shown in Fig.~\ref{fig:simu_setup}, the ambient plasma lies in the right half of the simulation box, while the left half is for the shocked plasma, flowing towards the right with an initial velocity of $v_1 = 1500$ km/s.
% (high-velocity case) or $v_1 = 500$ km/s (low-velocity case). 
Both of them consist of electrons and protons, with the real mass ratio $m_p/m_e=1836$.
The simulation box size is $L_x = 2048 d_e = 11$~mm, and the spatial resolution is $d_x = 0.2 d_e = 1.1$ $\mu$m, in which $d_e = c / \omega_{pe} = 5.3 \ \mu$m is the electron inertial length, and $\omega_{pe} = (n_{e0} q_e^2 / m_e / \epsilon_0)^{1/2} = 5.6 \times 10^{13}$ s$^{-1}$ is the electron plasma frequency. 
Here, c is the speed of light, $n_{e0} = 1.0 \times 10^{18}$ cm$^{-3}$ is the electron number density of the ambient plasma, and $m_e$, $q_e$ and $\epsilon_0$ are the electron mass, elementary charge, and the permittivity of free space, respectively. 
Note that the shock width is initialized to be equal to the ion inertial length $d_i = 200\ \mu$m.
The magnetic field is homogeneously applied in the z-direction with $B_z = 20$ T ($\omega_{ce} / \omega_{pe} = 0.06$, where $\omega_{ce} = q_e B / m_e$).
The simulation lasts for $1.5 \times 10^5 \omega_{pe}^{-1} \sim 2.5$~ns. 
Inside each cell, we put 1024 particles for each species. 
From the perspective of the ion Larmor motion, the simulation size is more than 10~$r_{Li}$, 
% and the simulation duration is more than 5~$\tau_{Li}$, 
in which $r_{Li} = v_1 / \omega_{ci} = m_i v_1 / q_e B \sim 0.8$~mm.
% and $\tau_{Li} = 1./\omega_{ci} \sim 0.5$~ns.

For the shocked plasma, the electron number density is $n_{e1} = 2n_{e0} = 2.0 \times 10^{18}$ cm$^{-3}$, and the temperature is $T_{e1} = 100$ eV and $T_{i1} = 200$ eV, all inferred from the TS characterization \cite{yao2020laboratory}. 
The boundary conditions for both particles and fields are open, and enough room is left between the boundary and the shock, so that the boundary conditions do not affect the concerned physics.
Given the initial low temperature of the ambient plasma in the simulation ($T_{e0} = 50$ eV), the Debye length is small compared to the grid resolution $d_x$, i.e. $\lambda_{De} = (\epsilon_0 kT_{e0}/n_{e0} q_e^2)^{1/2} \approx 0.01d_e = 0.05 d_x$. However, we do run a series of simulations with different initial temperatures, showing that the energy conservation for those cases is limited around $0.05\%$ and the physical results are almost the same. The mean-free-path of the presented case is $\lambda_{mfp} \approx 1800 d_e$, which is larger than the interaction scale, further confirming that the shock is collisionless.
% We have also tested a series of 1D and 2D simulations with varying resolutions ($0.1 \sim 0.5d_e$), particle-per-cell numbers ($512 \sim 2048$), ambient plasma temperatures ($20 \sim 100$ eV), and ion-to-mass ratio ($400 \sim 1836$) -- all reach similar shock dynamics as will be detailed in the following, which proves the robustness of the quasi-perpendicular, super-critical magnetized collisionless shock acceleration investigated here.

% \subsection{Simulation results}

\begin{figure*}[htp]
    \centering
    \includegraphics[width=0.6\textwidth]{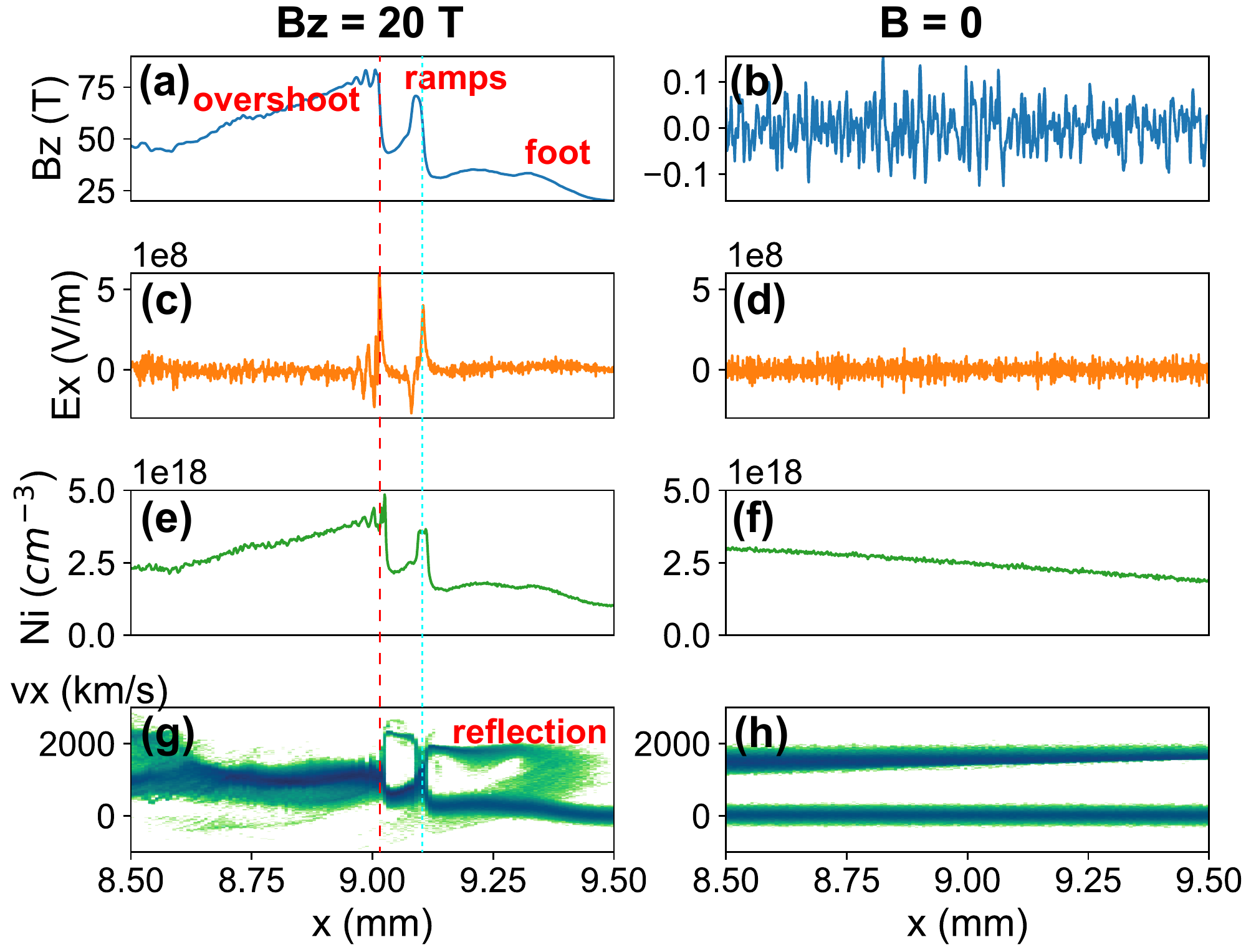}
    \caption{\textbf{Features of the super-critical quasi-perpendicular collisionless shock structure in ion density and EM fields distribution (with and without the external magnetic field), which prove the dominant particle acceleration mechanism to be SSA.} Specifically, (a) and (b) transverse magnetic field $B_z$; (c) and (d) longitudinal electric field $E_x$; (e) and (f) ion density profile; (g) and (h) phase-space distribution $x-v_x$ at the end of the simulation, i.e. at $t=2.7$ ns. The case with B-field is on the left column, while that without is on the right. The red dashed line and the cyan dotted line indicate the position of the shock ramps.}
    \label{fig:PIC_results}
\end{figure*}

\begin{figure*}[htp]
    \centering
    \includegraphics[width=0.6\textwidth]{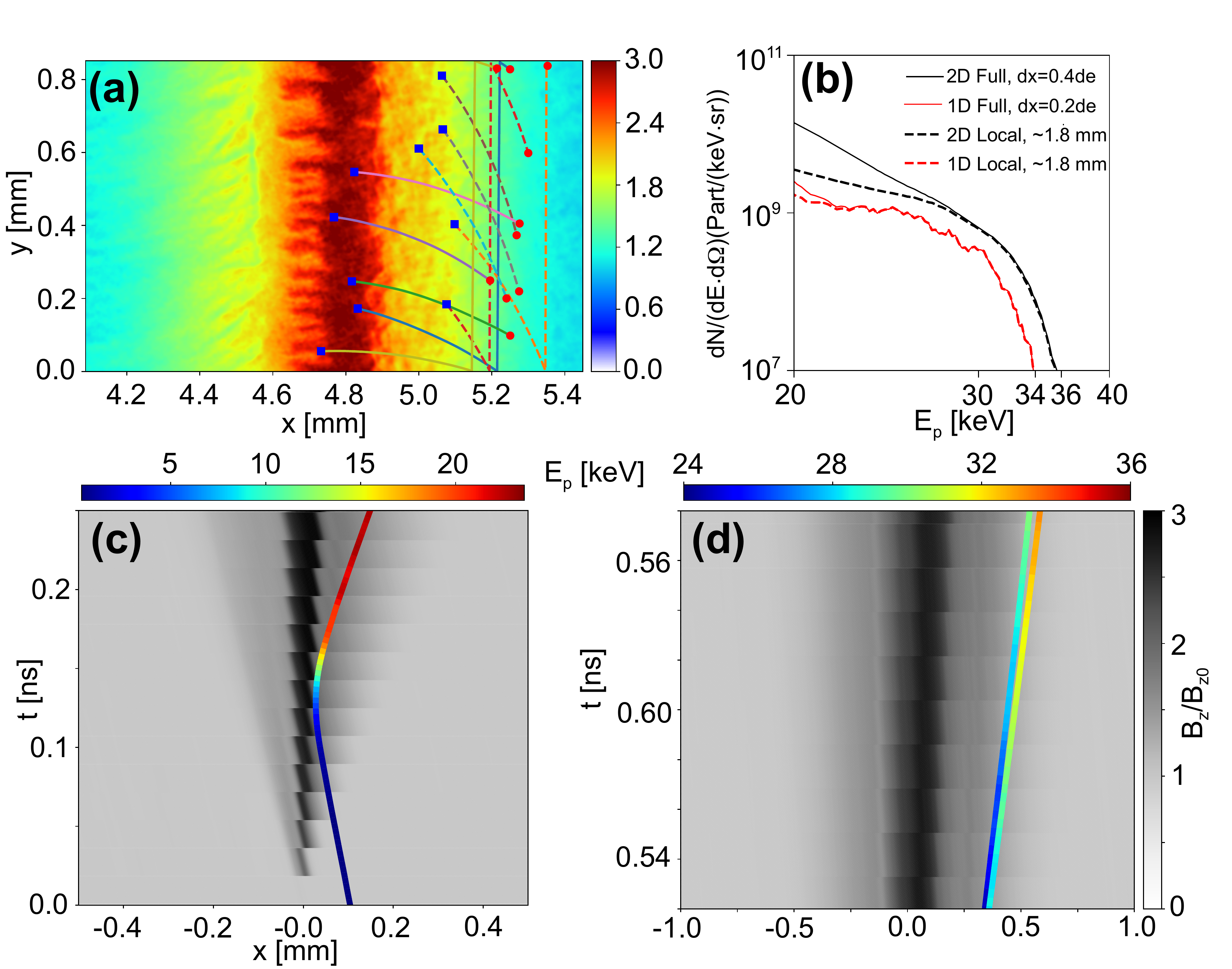}
    \caption{\textbf{2D simulation results.} (a) B-field maps at 0.7 ns, normalized to 20 T, with trajectories of protons ($E_k > 30$ keV). Solid lines are for protons from the ambient plasma and dashed ones are for protons from the drifting plasma; blue squares are the starting position at 0.5 ns, while red dots are the ending position at 0.7 ns. (b) Energy spectra of both 1D and 2D simulation results at 0.7 ns. Red lines are for the 1D case (solid line for protons in the whole simulation box, dashed line for those which lie around the shock layer in the vicinity of 1.8 mm), while black lines are for the corresponding 2D one. (c) Trajectory of a proton reflected at the shock front in the $x-t$ diagram, overlaid on the transversely-averaged B-field map in the reference frame of the contact discontinuity (the grey colorbar is for the B-field strength, while the colored one is for the proton kinetic energy). (d) Trajectories of two protons surfing along the shock front, also in the $x-t$ diagram, overlaid on the transversely-averaged B-field map in the same reference frame.
}
    \label{fig:2DPIC_results}
\end{figure*}

We report in Fig.~\ref{fig:PIC_results} the results of two PIC simulations, i.e. with and without B-fields.
For the case with the applied B-field (on the left column), typical structures of a super-critical quasi-perpendicular collisionless shock can be seen \cite{balogh2013physics}. 
For example, the overshoot in the DS region (on the left of the red dashed line), the ramps in the shock fronts (both the red dashed line and the cyan dotted line), and the foot in the upstream (US) region, as can be seen in Fig.~\ref{fig:PIC_results} (a). 
This foot region is formed by the reflected protons at a distance within $r_{L,i}$ and modulated by the modified two-stream instability \cite{matsukiyo2003modified}.
The proton density ($n_i$) in Fig.~\ref{fig:PIC_results} (e) shows a compression ratio of $n_{i,DS} / n_{i,US} \approx 4$, which agrees with the theoretical jump condition prediction \cite{woods1972shock}. 
This density profile, together with the transverse electric field $E_y$ (not shown here), also follows the distribution of the external applied B-field $B_z$. 
The longitudinal electric field ($E_x$) in Fig.~\ref{fig:PIC_results} (c) peaks right at the ramps, providing the electrostatic cross-shock potential to trap and reflect the protons, as can be seen in the phase-space distribution in Fig.~\ref{fig:PIC_results} (g).
Because the proton reflection is clearly due to the $E_x$ in our case, not the DS compressed B-field \cite{zank1996interstellar}, together with the fact that the ion Larmor radius (about 0.8 mm) is larger than the shock width (around 200 $\mu$m), the dominant particle acceleration mechanism is SSA, not SDA.
% This rules out the possibility of SDA being the dominant particle acceleration mechanism, because the ion reflection would be caused by the downstream compressed B-field \cite{zank1996interstellar}, not the longitudinal $E_x$ here in our case.
Note that the cyan dotted line indicates one of the periodic shock reformation \cite{balogh2013physics}.
On the contrary, for the case without B-field (on the right column), the drifting plasma just penetrates through the ambient gas and no shock is formed, thus no proton energization can take place, which is in accordance with our experimental observation. 

Note that in Fig.~\ref{fig:PIC_results} (c), the PIC simulation gives a longitudinal electric field $E_x \sim 5 \times 10^{8}$ V/m in the shock layer, which is two order-of-magnitude higher than the fitting of the proton radiography in Fig.~\ref{fig:ProtonRadio} (b). This discrepancy may be due to two reasons: on the one hand, the bipolar electric field structure fitted in the proton radiography has a size of 0.4 mm, while the $E_x$ peaks in the PIC simulations are very sharp, with their width smaller than 0.02 mm. With a time-average of the PIC simulation over 0.2 ns, the $E_x$ profile around the shock front reaches a size of 0.4 mm, and its value drops down to $2\times 10^7$ V/m. On the other hand, the PIC simulation represents the tip of the semi-sphere shaped expanding shock front at a single slice of z-direction, where the B-field is strictly perpendicular to the plasma flow and the shock is the strongest; however, the proton radiography covers the whole shock front with an integration along the z-direction. It includes all other plasma flow directions in the xz-plane, which are not perpendicular to the B-field and the corresponding shocks are weaker. Together with the above two aspects of reasons, it is understandable that the electric field fitted from the proton radiography shall be smaller than the PIC result. 

Particle dynamics of a high-velocity shock (as well as the comparison with the low-velocity case) and of the subsequent shock surfing proton energization is detailed in our previous paper \cite{yao2020laboratory}, while here we focus on demonstrating the robustness of the SSA mechanism that is at play in our experiment via 2D simulations, taking the non-stationarity \cite{burgess2007shock} into consideration.
Due to the limitation of the computational resources, we reduce the 2D simulation scale to an acceptable level: the simulation box size are $L_x = 8$ mm, $L_y = 0.8$ mm, the simulation time $t_{end} = 0.7$ ns, and the resolution is $d_x = 0.4 d_e$. 
% Specifically, with $r_{Li} = 0.8$ mm and $\tau_{Li} = 0.5$ ns, the 2D simulation time of more than $\tau_{Li}$ and transverse size of more than $r_{Li}$ should be enough to include the non-stationarity \cite{burgess2007shock} in 2D.

From Fig.~\ref{fig:2DPIC_results} (a), we can clearly see that the transverse non-stationarity has already occurred, with 2D-stripes mostly positioned at/behind the shock layer; while for protons with kinetic energy above 30 keV, their trajectories show that they mainly appear at the shock front, travelling down the negative y-direction. Note that the convective electric field $\bm{E} = - \bm{v} \times \bm{B}$ is towards the positive y-direction, i.e. $E_y = v_x B_z$; and the drifting of the protons against the convective electric field serves as a distinctive feature that the dominate proton acceleration mechanism is SSA, not SDA \cite{guo2013acceleration}.
% which further confirms that the acceleration is due to the convective electric field $E_y = v_x B_z$. 
Fig.~\ref{fig:2DPIC_results} (b) shows the proton energy spectra at 0.7 ns of both the 1D and 2D cases, which are close to each other, and there is only a 2 eV difference in the highest energy cut, which can be caused by the numerical heating of the 2D case (with lower spatial resolution). Moreover, checking the energy evolution of the protons in the $x-t$ diagram, overlaid on the transversely-averaged B-field map in the reference frame of the contact discontinuity (CD), it is clearly demonstrated that the accelerated proton is first reflected at (or, picked up by) the shock front in Fig.~\ref{fig:2DPIC_results} (c), and then surfing along the shock front while keeping gaining energy in Fig.~\ref{fig:2DPIC_results} (d). This is exactly the same picture as we have shown for the 1D simulations \cite{yao2020laboratory}, proving that the SSA is the dominating proton acceleration mechanism at play (even in the multi-dimensional case).

Nevertheless, the non-stationarity of the shock might further accelerate the proton at a later time, especially after the protons pass through the shock front and gyrate in the DS region. But unfortunately right now we do not have the computational resources to reveal that scenario. 
In short, our simulation shows that the non-stationarity does not prevent the protons being accelerated by SSA (reflecting and surfing), at least not at an early time.

\section{Conclusions}\label{conclusion}

In conclusion, we have shown that laboratory experiments can be performed to generate and characterize globally mildly super-critical, quasi-perpendicular magnetized collisionless shocks.
More importantly, non-thermal proton spectra are observed for the first time, and the underlying acceleration mechanism is pinpointed to be SSA via kinetic simulations, which can remarkably reproduce the experimental proton spectra.
Such laboratory studies for proton acceleration, as well as those for electrons reviewed above, can not only further our understanding of the shock formation and evolution by complementing spacecraft and remote sensing observations, but also help shed new light on solving the fundamental issue of injection for the UHECR production.

Our platform can be tuned in the future to perform a systematical study of collisionless shock with different B-field strength and orientation, enabling us to capture the transition of the magnetized collisionless shock from sub-critical regime to super-critical one, so that we can explore the triggering of the other acceleration scenarios (e.g. SDA and DSA).

\begin{acknowledgments}
The authors would like to thank the teams of the LULI (France) and JLF laser (USA) facilities for their expert support, as well the Dresden High Magnetic Field Laboratory at Helmholtz-Zentrum Dresden-Rossendorf for the development of the pulsed power generator used at LULI. We thank the Smilei dev-team for technical support. We also thank Ph. Savoini (Sorbonne U., France), L. Gremillet and C. Ruyer (CEA-France) for discussions. W.Y. would like to thank R. Li (SZTU, China) for discussions. This work was supported by funding from the European Research Council (ERC) under the European Unions Horizon 2020 research and innovation program (Grant Agreement No. 787539). The computational resources of this work were supported by the National Sciences and Engineering Research Council of Canada (NSERC) and Compute Canada (Job: pve-323-ac). Part of the experimental system is covered by a patent (1000183285, 2013, INPI-France). The FLASH software used was developed, in part, by the DOE NNSA ASC- and the DOE Office of Science ASCR-supported Flash Center for Computational Science at the University of Chicago. We thank J. L. Dubois for providing us EOS and opacities. The research leading to these results is supported by Extreme Light Infrastructure Nuclear Physics (ELI- NP) Phase II, a project co-financed by the Romanian Government and European Union through the European Regional Development Fund, and by the project $\ ELI-RO-2020-23$ funded by IFA (Romania). JIHT RAS team members are supported by The Ministry of Science and Higher Education of the Russian Federation (Agreement with Joint Institute for High Temperatures RAS No 075-15-2020-785). The reported study was funded by the Russian Foundation for Basic Research, project No. 19-32-60008.
\end{acknowledgments}

% \appendix

% \section{Appendixes}

% \nocite{*}
\bibliographystyle{apsrev4-2}
\bibliography{long}% Produces the bibliography via BibTeX.

\begin{thebibliography}{64}%
\makeatletter
\providecommand \@ifxundefined [1]{%
 \@ifx{#1\undefined}
}%
\providecommand \@ifnum [1]{%
 \ifnum #1\expandafter \@firstoftwo
 \else \expandafter \@secondoftwo
 \fi
}%
\providecommand \@ifx [1]{%
 \ifx #1\expandafter \@firstoftwo
 \else \expandafter \@secondoftwo
 \fi
}%
\providecommand \natexlab [1]{#1}%
\providecommand \enquote  [1]{``#1''}%
\providecommand \bibnamefont  [1]{#1}%
\providecommand \bibfnamefont [1]{#1}%
\providecommand \citenamefont [1]{#1}%
\providecommand \href@noop [0]{\@secondoftwo}%
\providecommand \href [0]{\begingroup \@sanitize@url \@href}%
\providecommand \@href[1]{\@@startlink{#1}\@@href}%
\providecommand \@@href[1]{\endgroup#1\@@endlink}%
\providecommand \@sanitize@url [0]{\catcode `\\12\catcode `\$12\catcode
  `\&12\catcode `\#12\catcode `\^12\catcode `\_12\catcode `\%12\relax}%
\providecommand \@@startlink[1]{}%
\providecommand \@@endlink[0]{}%
\providecommand \url  [0]{\begingroup\@sanitize@url \@url }%
\providecommand \@url [1]{\endgroup\@href {#1}{\urlprefix }}%
\providecommand \urlprefix  [0]{URL }%
\providecommand \Eprint [0]{\href }%
\providecommand \doibase [0]{https://doi.org/}%
\providecommand \selectlanguage [0]{\@gobble}%
\providecommand \bibinfo  [0]{\@secondoftwo}%
\providecommand \bibfield  [0]{\@secondoftwo}%
\providecommand \translation [1]{[#1]}%
\providecommand \BibitemOpen [0]{}%
\providecommand \bibitemStop [0]{}%
\providecommand \bibitemNoStop [0]{.\EOS\space}%
\providecommand \EOS [0]{\spacefactor3000\relax}%
\providecommand \BibitemShut  [1]{\csname bibitem#1\endcsname}%
\let\auto@bib@innerbib\@empty
%</preamble>
\bibitem [{\citenamefont {Yao}\ \emph {et~al.}(2020)\citenamefont {Yao},
  \citenamefont {Fazzini}, \citenamefont {Chen}, \citenamefont {Burdonov},
  \citenamefont {Antici}, \citenamefont {B{\'e}ard}, \citenamefont
  {Bola{\~n}os}, \citenamefont {Ciardi}, \citenamefont {Diab}, \citenamefont
  {Filippov} \emph {et~al.}}]{yao2020laboratory}%
  \BibitemOpen
  \bibfield  {author} {\bibinfo {author} {\bibfnamefont {W.}~\bibnamefont
  {Yao}}, \bibinfo {author} {\bibfnamefont {A.}~\bibnamefont {Fazzini}},
  \bibinfo {author} {\bibfnamefont {S.}~\bibnamefont {Chen}}, \bibinfo {author}
  {\bibfnamefont {K.}~\bibnamefont {Burdonov}}, \bibinfo {author}
  {\bibfnamefont {P.}~\bibnamefont {Antici}}, \bibinfo {author} {\bibfnamefont
  {J.}~\bibnamefont {B{\'e}ard}}, \bibinfo {author} {\bibfnamefont
  {S.}~\bibnamefont {Bola{\~n}os}}, \bibinfo {author} {\bibfnamefont
  {A.}~\bibnamefont {Ciardi}}, \bibinfo {author} {\bibfnamefont
  {R.}~\bibnamefont {Diab}}, \bibinfo {author} {\bibfnamefont {E.}~\bibnamefont
  {Filippov}}, \emph {et~al.},\ }\href {https://arxiv.org/abs/2011.00135}
  {\bibfield  {journal} {\bibinfo  {journal} {arXiv preprint arXiv:2011.00135}\
  } (\bibinfo {year} {2020})}\BibitemShut {NoStop}%
\bibitem [{\citenamefont {{Helder}}\ \emph {et~al.}(2009)\citenamefont
  {{Helder}}, \citenamefont {{Vink}}, \citenamefont {{Bassa}}, \citenamefont
  {{Bamba}}, \citenamefont {{Bleeker}}, \citenamefont {{Funk}}, \citenamefont
  {{Ghavamian}}, \citenamefont {{van der Heyden}}, \citenamefont {{Verbunt}},\
  and\ \citenamefont {{Yamazaki}}}]{2009Sci...325..719H}%
  \BibitemOpen
  \bibfield  {author} {\bibinfo {author} {\bibfnamefont {E.~A.}\ \bibnamefont
  {{Helder}}}, \bibinfo {author} {\bibfnamefont {J.}~\bibnamefont {{Vink}}},
  \bibinfo {author} {\bibfnamefont {C.~G.}\ \bibnamefont {{Bassa}}}, \bibinfo
  {author} {\bibfnamefont {A.}~\bibnamefont {{Bamba}}}, \bibinfo {author}
  {\bibfnamefont {J.~A.~M.}\ \bibnamefont {{Bleeker}}}, \bibinfo {author}
  {\bibfnamefont {S.}~\bibnamefont {{Funk}}}, \bibinfo {author} {\bibfnamefont
  {P.}~\bibnamefont {{Ghavamian}}}, \bibinfo {author} {\bibfnamefont {K.~J.}\
  \bibnamefont {{van der Heyden}}}, \bibinfo {author} {\bibfnamefont
  {F.}~\bibnamefont {{Verbunt}}},\ and\ \bibinfo {author} {\bibfnamefont
  {R.}~\bibnamefont {{Yamazaki}}},\ }\href
  {https://doi.org/10.1126/science.1173383} {\bibfield  {journal} {\bibinfo
  {journal} {Science}\ }\textbf {\bibinfo {volume} {325}},\ \bibinfo {pages}
  {719} (\bibinfo {year} {2009})}\BibitemShut {NoStop}%
\bibitem [{\citenamefont {{Nikoli{\'c}}}\ \emph {et~al.}(2013)\citenamefont
  {{Nikoli{\'c}}}, \citenamefont {{van de Ven}}, \citenamefont {{Heng}},
  \citenamefont {{Kupko}}, \citenamefont {{Husemann}}, \citenamefont
  {{Raymond}}, \citenamefont {{Hughes}},\ and\ \citenamefont
  {{Falc{\'o}n-Barroso}}}]{2013Sci...340...45N}%
  \BibitemOpen
  \bibfield  {author} {\bibinfo {author} {\bibfnamefont {S.}~\bibnamefont
  {{Nikoli{\'c}}}}, \bibinfo {author} {\bibfnamefont {G.}~\bibnamefont {{van de
  Ven}}}, \bibinfo {author} {\bibfnamefont {K.}~\bibnamefont {{Heng}}},
  \bibinfo {author} {\bibfnamefont {D.}~\bibnamefont {{Kupko}}}, \bibinfo
  {author} {\bibfnamefont {B.}~\bibnamefont {{Husemann}}}, \bibinfo {author}
  {\bibfnamefont {J.~C.}\ \bibnamefont {{Raymond}}}, \bibinfo {author}
  {\bibfnamefont {J.~P.}\ \bibnamefont {{Hughes}}},\ and\ \bibinfo {author}
  {\bibfnamefont {J.}~\bibnamefont {{Falc{\'o}n-Barroso}}},\ }\href
  {https://doi.org/10.1126/science.1228297} {\bibfield  {journal} {\bibinfo
  {journal} {Science}\ }\textbf {\bibinfo {volume} {340}},\ \bibinfo {pages}
  {45} (\bibinfo {year} {2013})}\BibitemShut {NoStop}%
\bibitem [{\citenamefont {Turner}\ \emph {et~al.}(2018)\citenamefont {Turner},
  \citenamefont {Wilson~III}, \citenamefont {Cohen}, \citenamefont {Schwartz},
  \citenamefont {Osmane}, \citenamefont {Fennell}, \citenamefont {Clemmons},
  \citenamefont {Blake}, \citenamefont {Westlake}, \citenamefont {Mauk},
  \citenamefont {Jaynes}, \citenamefont {Leonard}, \citenamefont {Baker},
  \citenamefont {Strangeway}, \citenamefont {Russell}, \citenamefont
  {Gershman}, \citenamefont {Avanov}, \citenamefont {Giles}, \citenamefont
  {Torbert}, \citenamefont {Broll}, \citenamefont {Gomez}, \citenamefont {A.},\
  and\ \citenamefont {Burch}}]{turner}%
  \BibitemOpen
  \bibfield  {author} {\bibinfo {author} {\bibfnamefont {D.~L.}\ \bibnamefont
  {Turner}}, \bibinfo {author} {\bibfnamefont {T.~Z.}\ \bibnamefont
  {Wilson~III}, \bibfnamefont {L.~B.~Liu}}, \bibinfo {author} {\bibfnamefont
  {I.~J.}\ \bibnamefont {Cohen}}, \bibinfo {author} {\bibfnamefont {S.~J.}\
  \bibnamefont {Schwartz}}, \bibinfo {author} {\bibfnamefont {A.}~\bibnamefont
  {Osmane}}, \bibinfo {author} {\bibfnamefont {J.~F.}\ \bibnamefont {Fennell}},
  \bibinfo {author} {\bibfnamefont {J.~H.}\ \bibnamefont {Clemmons}}, \bibinfo
  {author} {\bibfnamefont {J.~B.}\ \bibnamefont {Blake}}, \bibinfo {author}
  {\bibfnamefont {J.}~\bibnamefont {Westlake}}, \bibinfo {author}
  {\bibfnamefont {B.~H.}\ \bibnamefont {Mauk}}, \bibinfo {author}
  {\bibfnamefont {A.~N.}\ \bibnamefont {Jaynes}}, \bibinfo {author}
  {\bibfnamefont {T.}~\bibnamefont {Leonard}}, \bibinfo {author} {\bibfnamefont
  {D.~N.}\ \bibnamefont {Baker}}, \bibinfo {author} {\bibfnamefont {R.~J.}\
  \bibnamefont {Strangeway}}, \bibinfo {author} {\bibfnamefont {C.~T.}\
  \bibnamefont {Russell}}, \bibinfo {author} {\bibfnamefont {D.~J.}\
  \bibnamefont {Gershman}}, \bibinfo {author} {\bibfnamefont {L.}~\bibnamefont
  {Avanov}}, \bibinfo {author} {\bibfnamefont {B.~L.}\ \bibnamefont {Giles}},
  \bibinfo {author} {\bibfnamefont {R.~B.}\ \bibnamefont {Torbert}}, \bibinfo
  {author} {\bibfnamefont {J.}~\bibnamefont {Broll}}, \bibinfo {author}
  {\bibfnamefont {R.~G.}\ \bibnamefont {Gomez}}, \bibinfo {author}
  {\bibfnamefont {F.~S.}\ \bibnamefont {A.}},\ and\ \bibinfo {author}
  {\bibfnamefont {J.~L.}\ \bibnamefont {Burch}},\ }\href@noop {} {\bibfield
  {journal} {\bibinfo  {journal} {Nature}\ }\textbf {\bibinfo {volume} {561}},\
  \bibinfo {pages} {206–210} (\bibinfo {year} {2018})}\BibitemShut {NoStop}%
\bibitem [{\citenamefont {Amano}\ \emph {et~al.}(2020)\citenamefont {Amano},
  \citenamefont {Katou}, \citenamefont {Kitamura}, \citenamefont {Oka},
  \citenamefont {Matsumoto}, \citenamefont {Hoshino}, \citenamefont {Saito},
  \citenamefont {Yokota}, \citenamefont {Giles}, \citenamefont {Paterson} \emph
  {et~al.}}]{amano2020observational}%
  \BibitemOpen
  \bibfield  {author} {\bibinfo {author} {\bibfnamefont {T.}~\bibnamefont
  {Amano}}, \bibinfo {author} {\bibfnamefont {T.}~\bibnamefont {Katou}},
  \bibinfo {author} {\bibfnamefont {N.}~\bibnamefont {Kitamura}}, \bibinfo
  {author} {\bibfnamefont {M.}~\bibnamefont {Oka}}, \bibinfo {author}
  {\bibfnamefont {Y.}~\bibnamefont {Matsumoto}}, \bibinfo {author}
  {\bibfnamefont {M.}~\bibnamefont {Hoshino}}, \bibinfo {author} {\bibfnamefont
  {Y.}~\bibnamefont {Saito}}, \bibinfo {author} {\bibfnamefont
  {S.}~\bibnamefont {Yokota}}, \bibinfo {author} {\bibfnamefont
  {B.}~\bibnamefont {Giles}}, \bibinfo {author} {\bibfnamefont
  {W.}~\bibnamefont {Paterson}}, \emph {et~al.},\ }\href@noop {} {\bibfield
  {journal} {\bibinfo  {journal} {Physical Review Letters}\ }\textbf {\bibinfo
  {volume} {124}},\ \bibinfo {pages} {065101} (\bibinfo {year}
  {2020})}\BibitemShut {NoStop}%
\bibitem [{\citenamefont {Decker}\ \emph {et~al.}(2008)\citenamefont {Decker},
  \citenamefont {Krimigis}, \citenamefont {Roelof}, \citenamefont {Hill},
  \citenamefont {Armstrong}, \citenamefont {Gloeckler}, \citenamefont
  {Hamilton},\ and\ \citenamefont {Lanzerotti}}]{decker2008voyager2}%
  \BibitemOpen
  \bibfield  {author} {\bibinfo {author} {\bibfnamefont {R.}~\bibnamefont
  {Decker}}, \bibinfo {author} {\bibfnamefont {S.}~\bibnamefont {Krimigis}},
  \bibinfo {author} {\bibfnamefont {E.}~\bibnamefont {Roelof}}, \bibinfo
  {author} {\bibfnamefont {M.}~\bibnamefont {Hill}}, \bibinfo {author}
  {\bibfnamefont {T.}~\bibnamefont {Armstrong}}, \bibinfo {author}
  {\bibfnamefont {G.}~\bibnamefont {Gloeckler}}, \bibinfo {author}
  {\bibfnamefont {D.}~\bibnamefont {Hamilton}},\ and\ \bibinfo {author}
  {\bibfnamefont {L.}~\bibnamefont {Lanzerotti}},\ }\href@noop {} {\bibfield
  {journal} {\bibinfo  {journal} {Nature}\ }\textbf {\bibinfo {volume} {454}},\
  \bibinfo {pages} {67–70} (\bibinfo {year} {2008})}\BibitemShut {NoStop}%
\bibitem [{\citenamefont {Coroniti}(1970)}]{coroniti1970dissipation}%
  \BibitemOpen
  \bibfield  {author} {\bibinfo {author} {\bibfnamefont {F.}~\bibnamefont
  {Coroniti}},\ }\href@noop {} {\bibfield  {journal} {\bibinfo  {journal}
  {Journal of Plasma Physics}\ }\textbf {\bibinfo {volume} {4}},\ \bibinfo
  {pages} {265} (\bibinfo {year} {1970})}\BibitemShut {NoStop}%
\bibitem [{\citenamefont {Edmiston}\ and\ \citenamefont
  {Kennel}(1984)}]{edmiston1984parametric}%
  \BibitemOpen
  \bibfield  {author} {\bibinfo {author} {\bibfnamefont {J.}~\bibnamefont
  {Edmiston}}\ and\ \bibinfo {author} {\bibfnamefont {C.}~\bibnamefont
  {Kennel}},\ }\href@noop {} {\bibfield  {journal} {\bibinfo  {journal}
  {Journal of Plasma Physics}\ }\textbf {\bibinfo {volume} {32}},\ \bibinfo
  {pages} {429} (\bibinfo {year} {1984})}\BibitemShut {NoStop}%
\bibitem [{\citenamefont {Balogh}\ and\ \citenamefont
  {Treumann}(2013)}]{balogh2013physics}%
  \BibitemOpen
  \bibfield  {author} {\bibinfo {author} {\bibfnamefont {A.}~\bibnamefont
  {Balogh}}\ and\ \bibinfo {author} {\bibfnamefont {R.~A.}\ \bibnamefont
  {Treumann}},\ }\href@noop {} {\emph {\bibinfo {title} {Physics of
  collisionless shocks: space plasma shock waves}}}\ (\bibinfo  {publisher}
  {Springer New York},\ \bibinfo {address} {New York, NY},\ \bibinfo {year}
  {2013})\BibitemShut {NoStop}%
\bibitem [{\citenamefont {Zank}\ \emph {et~al.}(1996)\citenamefont {Zank},
  \citenamefont {Pauls}, \citenamefont {Cairns},\ and\ \citenamefont
  {Webb}}]{zank1996interstellar}%
  \BibitemOpen
  \bibfield  {author} {\bibinfo {author} {\bibfnamefont {G.}~\bibnamefont
  {Zank}}, \bibinfo {author} {\bibfnamefont {H.}~\bibnamefont {Pauls}},
  \bibinfo {author} {\bibfnamefont {I.}~\bibnamefont {Cairns}},\ and\ \bibinfo
  {author} {\bibfnamefont {G.}~\bibnamefont {Webb}},\ }\href@noop {} {\bibfield
   {journal} {\bibinfo  {journal} {Journal of Geophysical Research: Space
  Physics}\ }\textbf {\bibinfo {volume} {101}},\ \bibinfo {pages} {457}
  (\bibinfo {year} {1996})}\BibitemShut {NoStop}%
\bibitem [{\citenamefont {Lemb{\`e}ge}\ \emph {et~al.}(2004)\citenamefont
  {Lemb{\`e}ge}, \citenamefont {Giacalone}, \citenamefont {Scholer},
  \citenamefont {Hada}, \citenamefont {Hoshino}, \citenamefont
  {Krasnoselskikh}, \citenamefont {Kucharek}, \citenamefont {Savoini},\ and\
  \citenamefont {Terasawa}}]{lembege2004selected}%
  \BibitemOpen
  \bibfield  {author} {\bibinfo {author} {\bibfnamefont {B.}~\bibnamefont
  {Lemb{\`e}ge}}, \bibinfo {author} {\bibfnamefont {J.}~\bibnamefont
  {Giacalone}}, \bibinfo {author} {\bibfnamefont {M.}~\bibnamefont {Scholer}},
  \bibinfo {author} {\bibfnamefont {T.}~\bibnamefont {Hada}}, \bibinfo {author}
  {\bibfnamefont {M.}~\bibnamefont {Hoshino}}, \bibinfo {author} {\bibfnamefont
  {V.}~\bibnamefont {Krasnoselskikh}}, \bibinfo {author} {\bibfnamefont
  {H.}~\bibnamefont {Kucharek}}, \bibinfo {author} {\bibfnamefont
  {P.}~\bibnamefont {Savoini}},\ and\ \bibinfo {author} {\bibfnamefont
  {T.}~\bibnamefont {Terasawa}},\ }\href@noop {} {\bibfield  {journal}
  {\bibinfo  {journal} {Space Science Reviews}\ }\textbf {\bibinfo {volume}
  {110}},\ \bibinfo {pages} {161} (\bibinfo {year} {2004})}\BibitemShut
  {NoStop}%
\bibitem [{\citenamefont {Burrows}\ \emph {et~al.}(2010)\citenamefont
  {Burrows}, \citenamefont {Zank}, \citenamefont {Webb}, \citenamefont
  {Burlaga},\ and\ \citenamefont {Ness}}]{burrows2010pickup}%
  \BibitemOpen
  \bibfield  {author} {\bibinfo {author} {\bibfnamefont {R.}~\bibnamefont
  {Burrows}}, \bibinfo {author} {\bibfnamefont {G.}~\bibnamefont {Zank}},
  \bibinfo {author} {\bibfnamefont {G.}~\bibnamefont {Webb}}, \bibinfo {author}
  {\bibfnamefont {L.}~\bibnamefont {Burlaga}},\ and\ \bibinfo {author}
  {\bibfnamefont {N.}~\bibnamefont {Ness}},\ }\href@noop {} {\bibfield
  {journal} {\bibinfo  {journal} {The Astrophysical Journal}\ }\textbf
  {\bibinfo {volume} {715}},\ \bibinfo {pages} {1109} (\bibinfo {year}
  {2010})}\BibitemShut {NoStop}%
\bibitem [{\citenamefont {Zank}\ \emph {et~al.}(2009)\citenamefont {Zank},
  \citenamefont {Heerikhuisen}, \citenamefont {Pogorelov}, \citenamefont
  {Burrows},\ and\ \citenamefont {McComas}}]{zank2009microstructure}%
  \BibitemOpen
  \bibfield  {author} {\bibinfo {author} {\bibfnamefont {G.}~\bibnamefont
  {Zank}}, \bibinfo {author} {\bibfnamefont {J.}~\bibnamefont {Heerikhuisen}},
  \bibinfo {author} {\bibfnamefont {N.}~\bibnamefont {Pogorelov}}, \bibinfo
  {author} {\bibfnamefont {R.}~\bibnamefont {Burrows}},\ and\ \bibinfo {author}
  {\bibfnamefont {D.}~\bibnamefont {McComas}},\ }\href@noop {} {\bibfield
  {journal} {\bibinfo  {journal} {The Astrophysical Journal}\ }\textbf
  {\bibinfo {volume} {708}},\ \bibinfo {pages} {1092} (\bibinfo {year}
  {2009})}\BibitemShut {NoStop}%
\bibitem [{\citenamefont {Chalov}\ \emph {et~al.}(2016)\citenamefont {Chalov},
  \citenamefont {Malama}, \citenamefont {Alexashov},\ and\ \citenamefont
  {Izmodenov}}]{chalov2016acceleration}%
  \BibitemOpen
  \bibfield  {author} {\bibinfo {author} {\bibfnamefont {S.}~\bibnamefont
  {Chalov}}, \bibinfo {author} {\bibfnamefont {Y.}~\bibnamefont {Malama}},
  \bibinfo {author} {\bibfnamefont {D.}~\bibnamefont {Alexashov}},\ and\
  \bibinfo {author} {\bibfnamefont {V.}~\bibnamefont {Izmodenov}},\ }\href@noop
  {} {\bibfield  {journal} {\bibinfo  {journal} {Monthly Notices of the Royal
  Astronomical Society}\ }\textbf {\bibinfo {volume} {455}},\ \bibinfo {pages}
  {431} (\bibinfo {year} {2016})}\BibitemShut {NoStop}%
\bibitem [{\citenamefont {Guo}\ and\ \citenamefont
  {Giacalone}(2013)}]{guo2013acceleration}%
  \BibitemOpen
  \bibfield  {author} {\bibinfo {author} {\bibfnamefont {F.}~\bibnamefont
  {Guo}}\ and\ \bibinfo {author} {\bibfnamefont {J.}~\bibnamefont
  {Giacalone}},\ }\href@noop {} {\bibfield  {journal} {\bibinfo  {journal} {The
  Astrophysical Journal}\ }\textbf {\bibinfo {volume} {773}},\ \bibinfo {pages}
  {158} (\bibinfo {year} {2013})}\BibitemShut {NoStop}%
\bibitem [{\citenamefont {Yang}\ \emph {et~al.}(2009)\citenamefont {Yang},
  \citenamefont {Lu}, \citenamefont {Lemb{\`e}ge},\ and\ \citenamefont
  {Wang}}]{yang2009shock}%
  \BibitemOpen
  \bibfield  {author} {\bibinfo {author} {\bibfnamefont {Z.}~\bibnamefont
  {Yang}}, \bibinfo {author} {\bibfnamefont {Q.}~\bibnamefont {Lu}}, \bibinfo
  {author} {\bibfnamefont {B.}~\bibnamefont {Lemb{\`e}ge}},\ and\ \bibinfo
  {author} {\bibfnamefont {S.}~\bibnamefont {Wang}},\ }\href@noop {} {\bibfield
   {journal} {\bibinfo  {journal} {Journal of Geophysical Research: Space
  Physics}\ }\textbf {\bibinfo {volume} {114}} (\bibinfo {year}
  {2009})}\BibitemShut {NoStop}%
\bibitem [{\citenamefont {Yang}\ \emph {et~al.}(2012)\citenamefont {Yang},
  \citenamefont {Lemb{\`e}ge},\ and\ \citenamefont {Lu}}]{yang2012impact}%
  \BibitemOpen
  \bibfield  {author} {\bibinfo {author} {\bibfnamefont {Z.}~\bibnamefont
  {Yang}}, \bibinfo {author} {\bibfnamefont {B.}~\bibnamefont {Lemb{\`e}ge}},\
  and\ \bibinfo {author} {\bibfnamefont {Q.}~\bibnamefont {Lu}},\ }\href@noop
  {} {\bibfield  {journal} {\bibinfo  {journal} {Journal of Geophysical
  Research: Space Physics}\ }\textbf {\bibinfo {volume} {117}} (\bibinfo {year}
  {2012})}\BibitemShut {NoStop}%
\bibitem [{\citenamefont {Paul~Drake}(2006)}]{paul2006high}%
  \BibitemOpen
  \bibfield  {author} {\bibinfo {author} {\bibfnamefont {R.}~\bibnamefont
  {Paul~Drake}},\ }\href@noop {} {\emph {\bibinfo {title} {High Energy Density
  Physics: Fundamentals, Inertial Fusion and Experimental Astrophysics}}}\
  (\bibinfo  {publisher} {Springer-Verlag Berlin Heidelberg},\ \bibinfo {year}
  {2006})\BibitemShut {NoStop}%
\bibitem [{\citenamefont {Lebedev}\ \emph {et~al.}(2019)\citenamefont
  {Lebedev}, \citenamefont {Frank},\ and\ \citenamefont
  {Ryutov}}]{lebedev2019exploring}%
  \BibitemOpen
  \bibfield  {author} {\bibinfo {author} {\bibfnamefont {S.}~\bibnamefont
  {Lebedev}}, \bibinfo {author} {\bibfnamefont {A.}~\bibnamefont {Frank}},\
  and\ \bibinfo {author} {\bibfnamefont {D.}~\bibnamefont {Ryutov}},\
  }\href@noop {} {\bibfield  {journal} {\bibinfo  {journal} {Reviews of Modern
  Physics}\ }\textbf {\bibinfo {volume} {91}},\ \bibinfo {pages} {025002}
  (\bibinfo {year} {2019})}\BibitemShut {NoStop}%
\bibitem [{\citenamefont {Fox}\ \emph {et~al.}(2013)\citenamefont {Fox},
  \citenamefont {Fiksel}, \citenamefont {Bhattacharjee}, \citenamefont {Chang},
  \citenamefont {Germaschewski}, \citenamefont {Hu},\ and\ \citenamefont
  {Nilson}}]{fox2013filamentation}%
  \BibitemOpen
  \bibfield  {author} {\bibinfo {author} {\bibfnamefont {W.}~\bibnamefont
  {Fox}}, \bibinfo {author} {\bibfnamefont {G.}~\bibnamefont {Fiksel}},
  \bibinfo {author} {\bibfnamefont {A.}~\bibnamefont {Bhattacharjee}}, \bibinfo
  {author} {\bibfnamefont {P.-Y.}\ \bibnamefont {Chang}}, \bibinfo {author}
  {\bibfnamefont {K.}~\bibnamefont {Germaschewski}}, \bibinfo {author}
  {\bibfnamefont {S.}~\bibnamefont {Hu}},\ and\ \bibinfo {author}
  {\bibfnamefont {P.}~\bibnamefont {Nilson}},\ }\href@noop {} {\bibfield
  {journal} {\bibinfo  {journal} {Physical Review Letters}\ }\textbf {\bibinfo
  {volume} {111}},\ \bibinfo {pages} {225002} (\bibinfo {year}
  {2013})}\BibitemShut {NoStop}%
\bibitem [{\citenamefont {Huntington}\ \emph {et~al.}(2015)\citenamefont
  {Huntington}, \citenamefont {Fiuza}, \citenamefont {Ross}, \citenamefont
  {Zylstra}, \citenamefont {Drake}, \citenamefont {Froula}, \citenamefont
  {Gregori}, \citenamefont {Kugland}, \citenamefont {Kuranz}, \citenamefont
  {Levy} \emph {et~al.}}]{huntington2015observation}%
  \BibitemOpen
  \bibfield  {author} {\bibinfo {author} {\bibfnamefont {C.}~\bibnamefont
  {Huntington}}, \bibinfo {author} {\bibfnamefont {F.}~\bibnamefont {Fiuza}},
  \bibinfo {author} {\bibfnamefont {J.}~\bibnamefont {Ross}}, \bibinfo {author}
  {\bibfnamefont {A.}~\bibnamefont {Zylstra}}, \bibinfo {author} {\bibfnamefont
  {R.}~\bibnamefont {Drake}}, \bibinfo {author} {\bibfnamefont
  {D.}~\bibnamefont {Froula}}, \bibinfo {author} {\bibfnamefont
  {G.}~\bibnamefont {Gregori}}, \bibinfo {author} {\bibfnamefont
  {N.}~\bibnamefont {Kugland}}, \bibinfo {author} {\bibfnamefont
  {C.}~\bibnamefont {Kuranz}}, \bibinfo {author} {\bibfnamefont
  {M.}~\bibnamefont {Levy}}, \emph {et~al.},\ }\href@noop {} {\bibfield
  {journal} {\bibinfo  {journal} {Nature Physics}\ }\textbf {\bibinfo {volume}
  {11}},\ \bibinfo {pages} {173} (\bibinfo {year} {2015})}\BibitemShut
  {NoStop}%
\bibitem [{\citenamefont {Park}\ \emph {et~al.}(2015)\citenamefont {Park},
  \citenamefont {Huntington}, \citenamefont {Fiuza}, \citenamefont {Drake},
  \citenamefont {Froula}, \citenamefont {Gregori}, \citenamefont {Koenig},
  \citenamefont {Kugland}, \citenamefont {Kuranz}, \citenamefont {Lamb} \emph
  {et~al.}}]{park2015collisionless}%
  \BibitemOpen
  \bibfield  {author} {\bibinfo {author} {\bibfnamefont {H.-S.}\ \bibnamefont
  {Park}}, \bibinfo {author} {\bibfnamefont {C.}~\bibnamefont {Huntington}},
  \bibinfo {author} {\bibfnamefont {F.}~\bibnamefont {Fiuza}}, \bibinfo
  {author} {\bibfnamefont {R.}~\bibnamefont {Drake}}, \bibinfo {author}
  {\bibfnamefont {D.}~\bibnamefont {Froula}}, \bibinfo {author} {\bibfnamefont
  {G.}~\bibnamefont {Gregori}}, \bibinfo {author} {\bibfnamefont
  {M.}~\bibnamefont {Koenig}}, \bibinfo {author} {\bibfnamefont
  {N.}~\bibnamefont {Kugland}}, \bibinfo {author} {\bibfnamefont
  {C.}~\bibnamefont {Kuranz}}, \bibinfo {author} {\bibfnamefont
  {D.}~\bibnamefont {Lamb}}, \emph {et~al.},\ }\href@noop {} {\bibfield
  {journal} {\bibinfo  {journal} {Physics of Plasmas}\ }\textbf {\bibinfo
  {volume} {22}},\ \bibinfo {pages} {056311} (\bibinfo {year}
  {2015})}\BibitemShut {NoStop}%
\bibitem [{\citenamefont {Park}\ \emph {et~al.}(2016)\citenamefont {Park},
  \citenamefont {Ross}, \citenamefont {Huntington}, \citenamefont {Fiuza},
  \citenamefont {Ryutov}, \citenamefont {Casey}, \citenamefont {Drake},
  \citenamefont {Fiksel}, \citenamefont {Froula}, \citenamefont {Gregori},
  \citenamefont {Kugland}, \citenamefont {Kuranz}, \citenamefont {Levy},
  \citenamefont {Li}, \citenamefont {Meinecke}, \citenamefont {Morita},
  \citenamefont {Petrasso}, \citenamefont {Plechaty}, \citenamefont
  {Remington}, \citenamefont {Sakawa}, \citenamefont {Spitkovsky},
  \citenamefont {Takabe},\ and\ \citenamefont {Zylstra}}]{park2016laboratory}%
  \BibitemOpen
  \bibfield  {author} {\bibinfo {author} {\bibfnamefont {H.-S.}\ \bibnamefont
  {Park}}, \bibinfo {author} {\bibfnamefont {J.~S.}\ \bibnamefont {Ross}},
  \bibinfo {author} {\bibfnamefont {C.~M.}\ \bibnamefont {Huntington}},
  \bibinfo {author} {\bibfnamefont {F.}~\bibnamefont {Fiuza}}, \bibinfo
  {author} {\bibfnamefont {D.}~\bibnamefont {Ryutov}}, \bibinfo {author}
  {\bibfnamefont {D.}~\bibnamefont {Casey}}, \bibinfo {author} {\bibfnamefont
  {R.~P.}\ \bibnamefont {Drake}}, \bibinfo {author} {\bibfnamefont
  {G.}~\bibnamefont {Fiksel}}, \bibinfo {author} {\bibfnamefont
  {D.}~\bibnamefont {Froula}}, \bibinfo {author} {\bibfnamefont
  {G.}~\bibnamefont {Gregori}}, \bibinfo {author} {\bibfnamefont {N.~L.}\
  \bibnamefont {Kugland}}, \bibinfo {author} {\bibfnamefont {C.}~\bibnamefont
  {Kuranz}}, \bibinfo {author} {\bibfnamefont {M.~C.}\ \bibnamefont {Levy}},
  \bibinfo {author} {\bibfnamefont {C.~K.}\ \bibnamefont {Li}}, \bibinfo
  {author} {\bibfnamefont {J.}~\bibnamefont {Meinecke}}, \bibinfo {author}
  {\bibfnamefont {T.}~\bibnamefont {Morita}}, \bibinfo {author} {\bibfnamefont
  {R.}~\bibnamefont {Petrasso}}, \bibinfo {author} {\bibfnamefont
  {C.}~\bibnamefont {Plechaty}}, \bibinfo {author} {\bibfnamefont
  {B.}~\bibnamefont {Remington}}, \bibinfo {author} {\bibfnamefont
  {Y.}~\bibnamefont {Sakawa}}, \bibinfo {author} {\bibfnamefont
  {A.}~\bibnamefont {Spitkovsky}}, \bibinfo {author} {\bibfnamefont
  {H.}~\bibnamefont {Takabe}},\ and\ \bibinfo {author} {\bibfnamefont {A.~B.}\
  \bibnamefont {Zylstra}},\ }\href
  {https://doi.org/10.1088/1742-6596/688/1/012084} {\bibfield  {journal}
  {\bibinfo  {journal} {Journal of Physics: Conference Series}\ }\textbf
  {\bibinfo {volume} {688}},\ \bibinfo {pages} {012084} (\bibinfo {year}
  {2016})}\BibitemShut {NoStop}%
\bibitem [{\citenamefont {Ross}\ \emph {et~al.}(2017)\citenamefont {Ross},
  \citenamefont {Higginson}, \citenamefont {Ryutov}, \citenamefont {Fiuza},
  \citenamefont {Hatarik}, \citenamefont {Huntington}, \citenamefont
  {Kalantar}, \citenamefont {Link}, \citenamefont {Pollock}, \citenamefont
  {Remington} \emph {et~al.}}]{ross2017transition}%
  \BibitemOpen
  \bibfield  {author} {\bibinfo {author} {\bibfnamefont {J.}~\bibnamefont
  {Ross}}, \bibinfo {author} {\bibfnamefont {D.}~\bibnamefont {Higginson}},
  \bibinfo {author} {\bibfnamefont {D.}~\bibnamefont {Ryutov}}, \bibinfo
  {author} {\bibfnamefont {F.}~\bibnamefont {Fiuza}}, \bibinfo {author}
  {\bibfnamefont {R.}~\bibnamefont {Hatarik}}, \bibinfo {author} {\bibfnamefont
  {C.}~\bibnamefont {Huntington}}, \bibinfo {author} {\bibfnamefont
  {D.}~\bibnamefont {Kalantar}}, \bibinfo {author} {\bibfnamefont
  {A.}~\bibnamefont {Link}}, \bibinfo {author} {\bibfnamefont {B.}~\bibnamefont
  {Pollock}}, \bibinfo {author} {\bibfnamefont {B.}~\bibnamefont {Remington}},
  \emph {et~al.},\ }\href@noop {} {\bibfield  {journal} {\bibinfo  {journal}
  {Physical Review Letters}\ }\textbf {\bibinfo {volume} {118}},\ \bibinfo
  {pages} {185003} (\bibinfo {year} {2017})}\BibitemShut {NoStop}%
\bibitem [{\citenamefont {Courtois}\ \emph {et~al.}(2004)\citenamefont
  {Courtois}, \citenamefont {Grundy}, \citenamefont {Ash}, \citenamefont
  {Chambers}, \citenamefont {Woolsey}, \citenamefont {Dendy},\ and\
  \citenamefont {McClements}}]{courtois2004experiment}%
  \BibitemOpen
  \bibfield  {author} {\bibinfo {author} {\bibfnamefont {C.}~\bibnamefont
  {Courtois}}, \bibinfo {author} {\bibfnamefont {R.}~\bibnamefont {Grundy}},
  \bibinfo {author} {\bibfnamefont {A.}~\bibnamefont {Ash}}, \bibinfo {author}
  {\bibfnamefont {D.}~\bibnamefont {Chambers}}, \bibinfo {author}
  {\bibfnamefont {N.}~\bibnamefont {Woolsey}}, \bibinfo {author} {\bibfnamefont
  {R.}~\bibnamefont {Dendy}},\ and\ \bibinfo {author} {\bibfnamefont
  {K.}~\bibnamefont {McClements}},\ }\href@noop {} {\bibfield  {journal}
  {\bibinfo  {journal} {Physics of Plasmas}\ }\textbf {\bibinfo {volume}
  {11}},\ \bibinfo {pages} {3386} (\bibinfo {year} {2004})}\BibitemShut
  {NoStop}%
\bibitem [{\citenamefont {Yuan}\ \emph {et~al.}(2018)\citenamefont {Yuan},
  \citenamefont {Wei}, \citenamefont {Liang}, \citenamefont {Wang},
  \citenamefont {Li}, \citenamefont {Zhang}, \citenamefont {Zhu}, \citenamefont
  {Zhao}, \citenamefont {Jiang}, \citenamefont {Han} \emph
  {et~al.}}]{yuan2018laboratory}%
  \BibitemOpen
  \bibfield  {author} {\bibinfo {author} {\bibfnamefont {D.}~\bibnamefont
  {Yuan}}, \bibinfo {author} {\bibfnamefont {H.}~\bibnamefont {Wei}}, \bibinfo
  {author} {\bibfnamefont {G.}~\bibnamefont {Liang}}, \bibinfo {author}
  {\bibfnamefont {F.}~\bibnamefont {Wang}}, \bibinfo {author} {\bibfnamefont
  {Y.}~\bibnamefont {Li}}, \bibinfo {author} {\bibfnamefont {Z.}~\bibnamefont
  {Zhang}}, \bibinfo {author} {\bibfnamefont {B.}~\bibnamefont {Zhu}}, \bibinfo
  {author} {\bibfnamefont {J.}~\bibnamefont {Zhao}}, \bibinfo {author}
  {\bibfnamefont {W.}~\bibnamefont {Jiang}}, \bibinfo {author} {\bibfnamefont
  {B.}~\bibnamefont {Han}}, \emph {et~al.},\ }\href@noop {} {\bibfield
  {journal} {\bibinfo  {journal} {High Power Laser Science and Engineering}\
  }\textbf {\bibinfo {volume} {6}} (\bibinfo {year} {2018})}\BibitemShut
  {NoStop}%
\bibitem [{\citenamefont {Kuramitsu}\ \emph {et~al.}(2011)\citenamefont
  {Kuramitsu}, \citenamefont {Sakawa}, \citenamefont {Morita}, \citenamefont
  {Gregory}, \citenamefont {Waugh}, \citenamefont {Dono}, \citenamefont {Aoki},
  \citenamefont {Tanji}, \citenamefont {Koenig}, \citenamefont {Woolsey} \emph
  {et~al.}}]{kuramitsu2011time}%
  \BibitemOpen
  \bibfield  {author} {\bibinfo {author} {\bibfnamefont {Y.}~\bibnamefont
  {Kuramitsu}}, \bibinfo {author} {\bibfnamefont {Y.}~\bibnamefont {Sakawa}},
  \bibinfo {author} {\bibfnamefont {T.}~\bibnamefont {Morita}}, \bibinfo
  {author} {\bibfnamefont {C.}~\bibnamefont {Gregory}}, \bibinfo {author}
  {\bibfnamefont {J.}~\bibnamefont {Waugh}}, \bibinfo {author} {\bibfnamefont
  {S.}~\bibnamefont {Dono}}, \bibinfo {author} {\bibfnamefont {H.}~\bibnamefont
  {Aoki}}, \bibinfo {author} {\bibfnamefont {H.}~\bibnamefont {Tanji}},
  \bibinfo {author} {\bibfnamefont {M.}~\bibnamefont {Koenig}}, \bibinfo
  {author} {\bibfnamefont {N.}~\bibnamefont {Woolsey}}, \emph {et~al.},\
  }\href@noop {} {\bibfield  {journal} {\bibinfo  {journal} {Physical Review
  Letters}\ }\textbf {\bibinfo {volume} {106}},\ \bibinfo {pages} {175002}
  (\bibinfo {year} {2011})}\BibitemShut {NoStop}%
\bibitem [{\citenamefont {Li}\ \emph {et~al.}(2019)\citenamefont {Li},
  \citenamefont {Tikhonchuk}, \citenamefont {Moreno}, \citenamefont {Sio},
  \citenamefont {D’Humi{\`e}res}, \citenamefont {Ribeyre}, \citenamefont
  {Korneev}, \citenamefont {Atzeni}, \citenamefont {Betti}, \citenamefont
  {Birkel} \emph {et~al.}}]{li2019collisionless}%
  \BibitemOpen
  \bibfield  {author} {\bibinfo {author} {\bibfnamefont {C.}~\bibnamefont
  {Li}}, \bibinfo {author} {\bibfnamefont {V.}~\bibnamefont {Tikhonchuk}},
  \bibinfo {author} {\bibfnamefont {Q.}~\bibnamefont {Moreno}}, \bibinfo
  {author} {\bibfnamefont {H.}~\bibnamefont {Sio}}, \bibinfo {author}
  {\bibfnamefont {E.}~\bibnamefont {D’Humi{\`e}res}}, \bibinfo {author}
  {\bibfnamefont {X.}~\bibnamefont {Ribeyre}}, \bibinfo {author} {\bibfnamefont
  {P.}~\bibnamefont {Korneev}}, \bibinfo {author} {\bibfnamefont
  {S.}~\bibnamefont {Atzeni}}, \bibinfo {author} {\bibfnamefont
  {R.}~\bibnamefont {Betti}}, \bibinfo {author} {\bibfnamefont
  {A.}~\bibnamefont {Birkel}}, \emph {et~al.},\ }\href@noop {} {\bibfield
  {journal} {\bibinfo  {journal} {Physical Review Letters}\ }\textbf {\bibinfo
  {volume} {123}},\ \bibinfo {pages} {055002} (\bibinfo {year}
  {2019})}\BibitemShut {NoStop}%
\bibitem [{\citenamefont {Swadling}\ \emph {et~al.}(2020)\citenamefont
  {Swadling}, \citenamefont {Bruulsema}, \citenamefont {Fiuza}, \citenamefont
  {Higginson}, \citenamefont {Huntington}, \citenamefont {Park}, \citenamefont
  {Pollock}, \citenamefont {Rozmus}, \citenamefont {Rinderknecht},
  \citenamefont {Katz} \emph {et~al.}}]{swadling2020measurement}%
  \BibitemOpen
  \bibfield  {author} {\bibinfo {author} {\bibfnamefont {G.}~\bibnamefont
  {Swadling}}, \bibinfo {author} {\bibfnamefont {C.}~\bibnamefont {Bruulsema}},
  \bibinfo {author} {\bibfnamefont {F.}~\bibnamefont {Fiuza}}, \bibinfo
  {author} {\bibfnamefont {D.}~\bibnamefont {Higginson}}, \bibinfo {author}
  {\bibfnamefont {C.}~\bibnamefont {Huntington}}, \bibinfo {author}
  {\bibfnamefont {H.}~\bibnamefont {Park}}, \bibinfo {author} {\bibfnamefont
  {B.}~\bibnamefont {Pollock}}, \bibinfo {author} {\bibfnamefont
  {W.}~\bibnamefont {Rozmus}}, \bibinfo {author} {\bibfnamefont
  {H.}~\bibnamefont {Rinderknecht}}, \bibinfo {author} {\bibfnamefont
  {J.}~\bibnamefont {Katz}}, \emph {et~al.},\ }\href@noop {} {\bibfield
  {journal} {\bibinfo  {journal} {Physical Review Letters}\ }\textbf {\bibinfo
  {volume} {124}},\ \bibinfo {pages} {215001} (\bibinfo {year}
  {2020})}\BibitemShut {NoStop}%
\bibitem [{\citenamefont {Fiuza}\ \emph {et~al.}(2020)\citenamefont {Fiuza},
  \citenamefont {Swadling}, \citenamefont {Grassi}, \citenamefont
  {Rinderknecht}, \citenamefont {Higginson}, \citenamefont {Ryutov},
  \citenamefont {Bruulsema}, \citenamefont {Drake}, \citenamefont {Funk},
  \citenamefont {Glenzer} \emph {et~al.}}]{fiuza_electron_2020}%
  \BibitemOpen
  \bibfield  {author} {\bibinfo {author} {\bibfnamefont {F.}~\bibnamefont
  {Fiuza}}, \bibinfo {author} {\bibfnamefont {G.}~\bibnamefont {Swadling}},
  \bibinfo {author} {\bibfnamefont {A.}~\bibnamefont {Grassi}}, \bibinfo
  {author} {\bibfnamefont {H.}~\bibnamefont {Rinderknecht}}, \bibinfo {author}
  {\bibfnamefont {D.}~\bibnamefont {Higginson}}, \bibinfo {author}
  {\bibfnamefont {D.}~\bibnamefont {Ryutov}}, \bibinfo {author} {\bibfnamefont
  {C.}~\bibnamefont {Bruulsema}}, \bibinfo {author} {\bibfnamefont
  {R.}~\bibnamefont {Drake}}, \bibinfo {author} {\bibfnamefont
  {S.}~\bibnamefont {Funk}}, \bibinfo {author} {\bibfnamefont {S.}~\bibnamefont
  {Glenzer}}, \emph {et~al.},\ }\href@noop {} {\bibfield  {journal} {\bibinfo
  {journal} {Nature Physics}\ }\textbf {\bibinfo {volume} {16}},\ \bibinfo
  {pages} {916} (\bibinfo {year} {2020})}\BibitemShut {NoStop}%
\bibitem [{\citenamefont {Schaeffer}\ \emph {et~al.}(2012)\citenamefont
  {Schaeffer}, \citenamefont {Everson}, \citenamefont {Winske}, \citenamefont
  {Constantin}, \citenamefont {Bondarenko}, \citenamefont {Morton},
  \citenamefont {Flippo}, \citenamefont {Montgomery}, \citenamefont
  {Gaillard},\ and\ \citenamefont {Niemann}}]{schaeffer2012generation}%
  \BibitemOpen
  \bibfield  {author} {\bibinfo {author} {\bibfnamefont {D.}~\bibnamefont
  {Schaeffer}}, \bibinfo {author} {\bibfnamefont {E.}~\bibnamefont {Everson}},
  \bibinfo {author} {\bibfnamefont {D.}~\bibnamefont {Winske}}, \bibinfo
  {author} {\bibfnamefont {C.}~\bibnamefont {Constantin}}, \bibinfo {author}
  {\bibfnamefont {A.}~\bibnamefont {Bondarenko}}, \bibinfo {author}
  {\bibfnamefont {L.}~\bibnamefont {Morton}}, \bibinfo {author} {\bibfnamefont
  {K.}~\bibnamefont {Flippo}}, \bibinfo {author} {\bibfnamefont
  {D.}~\bibnamefont {Montgomery}}, \bibinfo {author} {\bibfnamefont
  {S.}~\bibnamefont {Gaillard}},\ and\ \bibinfo {author} {\bibfnamefont
  {C.}~\bibnamefont {Niemann}},\ }\href@noop {} {\bibfield  {journal} {\bibinfo
   {journal} {Physics of Plasmas}\ }\textbf {\bibinfo {volume} {19}},\ \bibinfo
  {pages} {070702} (\bibinfo {year} {2012})}\BibitemShut {NoStop}%
\bibitem [{\citenamefont {Niemann}\ \emph {et~al.}(2014)\citenamefont
  {Niemann}, \citenamefont {Gekelman}, \citenamefont {Constantin},
  \citenamefont {Everson}, \citenamefont {Schaeffer}, \citenamefont
  {Bondarenko}, \citenamefont {Clark}, \citenamefont {Winske}, \citenamefont
  {Vincena}, \citenamefont {Van~Compernolle} \emph
  {et~al.}}]{niemann2014observation}%
  \BibitemOpen
  \bibfield  {author} {\bibinfo {author} {\bibfnamefont {C.}~\bibnamefont
  {Niemann}}, \bibinfo {author} {\bibfnamefont {W.}~\bibnamefont {Gekelman}},
  \bibinfo {author} {\bibfnamefont {C.}~\bibnamefont {Constantin}}, \bibinfo
  {author} {\bibfnamefont {E.}~\bibnamefont {Everson}}, \bibinfo {author}
  {\bibfnamefont {D.}~\bibnamefont {Schaeffer}}, \bibinfo {author}
  {\bibfnamefont {A.}~\bibnamefont {Bondarenko}}, \bibinfo {author}
  {\bibfnamefont {S.}~\bibnamefont {Clark}}, \bibinfo {author} {\bibfnamefont
  {D.}~\bibnamefont {Winske}}, \bibinfo {author} {\bibfnamefont
  {S.}~\bibnamefont {Vincena}}, \bibinfo {author} {\bibfnamefont
  {B.}~\bibnamefont {Van~Compernolle}}, \emph {et~al.},\ }\href@noop {}
  {\bibfield  {journal} {\bibinfo  {journal} {Geophysical Research Letters}\
  }\textbf {\bibinfo {volume} {41}},\ \bibinfo {pages} {7413} (\bibinfo {year}
  {2014})}\BibitemShut {NoStop}%
\bibitem [{\citenamefont {Schaeffer}\ \emph
  {et~al.}(2017{\natexlab{a}})\citenamefont {Schaeffer}, \citenamefont {Fox},
  \citenamefont {Haberberger}, \citenamefont {Fiksel}, \citenamefont
  {Bhattacharjee}, \citenamefont {Barnak}, \citenamefont {Hu},\ and\
  \citenamefont {Germaschewski}}]{schaeffer2017generation}%
  \BibitemOpen
  \bibfield  {author} {\bibinfo {author} {\bibfnamefont {D.}~\bibnamefont
  {Schaeffer}}, \bibinfo {author} {\bibfnamefont {W.}~\bibnamefont {Fox}},
  \bibinfo {author} {\bibfnamefont {D.}~\bibnamefont {Haberberger}}, \bibinfo
  {author} {\bibfnamefont {G.}~\bibnamefont {Fiksel}}, \bibinfo {author}
  {\bibfnamefont {A.}~\bibnamefont {Bhattacharjee}}, \bibinfo {author}
  {\bibfnamefont {D.}~\bibnamefont {Barnak}}, \bibinfo {author} {\bibfnamefont
  {S.}~\bibnamefont {Hu}},\ and\ \bibinfo {author} {\bibfnamefont
  {K.}~\bibnamefont {Germaschewski}},\ }\href@noop {} {\bibfield  {journal}
  {\bibinfo  {journal} {Physical Review Letters}\ }\textbf {\bibinfo {volume}
  {119}},\ \bibinfo {pages} {025001} (\bibinfo {year}
  {2017}{\natexlab{a}})}\BibitemShut {NoStop}%
\bibitem [{\citenamefont {Schaeffer}\ \emph
  {et~al.}(2017{\natexlab{b}})\citenamefont {Schaeffer}, \citenamefont {Fox},
  \citenamefont {Haberberger}, \citenamefont {Fiksel}, \citenamefont
  {Bhattacharjee}, \citenamefont {Barnak}, \citenamefont {Hu}, \citenamefont
  {Germaschewski},\ and\ \citenamefont {Follett}}]{schaeffer2017high}%
  \BibitemOpen
  \bibfield  {author} {\bibinfo {author} {\bibfnamefont {D.~B.}\ \bibnamefont
  {Schaeffer}}, \bibinfo {author} {\bibfnamefont {W.}~\bibnamefont {Fox}},
  \bibinfo {author} {\bibfnamefont {D.}~\bibnamefont {Haberberger}}, \bibinfo
  {author} {\bibfnamefont {G.}~\bibnamefont {Fiksel}}, \bibinfo {author}
  {\bibfnamefont {A.}~\bibnamefont {Bhattacharjee}}, \bibinfo {author}
  {\bibfnamefont {D.}~\bibnamefont {Barnak}}, \bibinfo {author} {\bibfnamefont
  {S.}~\bibnamefont {Hu}}, \bibinfo {author} {\bibfnamefont {K.}~\bibnamefont
  {Germaschewski}},\ and\ \bibinfo {author} {\bibfnamefont {R.}~\bibnamefont
  {Follett}},\ }\href@noop {} {\bibfield  {journal} {\bibinfo  {journal}
  {Physics of Plasmas}\ }\textbf {\bibinfo {volume} {24}},\ \bibinfo {pages}
  {122702} (\bibinfo {year} {2017}{\natexlab{b}})}\BibitemShut {NoStop}%
\bibitem [{\citenamefont {Schaeffer}\ \emph {et~al.}(2019)\citenamefont
  {Schaeffer}, \citenamefont {Fox}, \citenamefont {Follett}, \citenamefont
  {Fiksel}, \citenamefont {Li}, \citenamefont {Matteucci}, \citenamefont
  {Bhattacharjee},\ and\ \citenamefont {Germaschewski}}]{schaeffer2019direct}%
  \BibitemOpen
  \bibfield  {author} {\bibinfo {author} {\bibfnamefont {D.~B.}\ \bibnamefont
  {Schaeffer}}, \bibinfo {author} {\bibfnamefont {W.}~\bibnamefont {Fox}},
  \bibinfo {author} {\bibfnamefont {R.}~\bibnamefont {Follett}}, \bibinfo
  {author} {\bibfnamefont {G.}~\bibnamefont {Fiksel}}, \bibinfo {author}
  {\bibfnamefont {C.}~\bibnamefont {Li}}, \bibinfo {author} {\bibfnamefont
  {J.}~\bibnamefont {Matteucci}}, \bibinfo {author} {\bibfnamefont
  {A.}~\bibnamefont {Bhattacharjee}},\ and\ \bibinfo {author} {\bibfnamefont
  {K.}~\bibnamefont {Germaschewski}},\ }\href@noop {} {\bibfield  {journal}
  {\bibinfo  {journal} {Physical Review Letters}\ }\textbf {\bibinfo {volume}
  {122}},\ \bibinfo {pages} {245001} (\bibinfo {year} {2019})}\BibitemShut
  {NoStop}%
\bibitem [{\citenamefont {Romagnani}\ \emph {et~al.}(2008)\citenamefont
  {Romagnani}, \citenamefont {Bulanov}, \citenamefont {Borghesi}, \citenamefont
  {Audebert}, \citenamefont {Gauthier}, \citenamefont {L{\"o}wenbr{\"u}ck},
  \citenamefont {Mackinnon}, \citenamefont {Patel}, \citenamefont {Pretzler},
  \citenamefont {Toncian} \emph {et~al.}}]{romagnani2008observation}%
  \BibitemOpen
  \bibfield  {author} {\bibinfo {author} {\bibfnamefont {L.}~\bibnamefont
  {Romagnani}}, \bibinfo {author} {\bibfnamefont {S.}~\bibnamefont {Bulanov}},
  \bibinfo {author} {\bibfnamefont {M.}~\bibnamefont {Borghesi}}, \bibinfo
  {author} {\bibfnamefont {P.}~\bibnamefont {Audebert}}, \bibinfo {author}
  {\bibfnamefont {J.}~\bibnamefont {Gauthier}}, \bibinfo {author}
  {\bibfnamefont {K.}~\bibnamefont {L{\"o}wenbr{\"u}ck}}, \bibinfo {author}
  {\bibfnamefont {A.}~\bibnamefont {Mackinnon}}, \bibinfo {author}
  {\bibfnamefont {P.}~\bibnamefont {Patel}}, \bibinfo {author} {\bibfnamefont
  {G.}~\bibnamefont {Pretzler}}, \bibinfo {author} {\bibfnamefont
  {T.}~\bibnamefont {Toncian}}, \emph {et~al.},\ }\href@noop {} {\bibfield
  {journal} {\bibinfo  {journal} {Physical Review Letters}\ }\textbf {\bibinfo
  {volume} {101}},\ \bibinfo {pages} {025004} (\bibinfo {year}
  {2008})}\BibitemShut {NoStop}%
\bibitem [{\citenamefont {Rigby}\ \emph {et~al.}(2018)\citenamefont {Rigby},
  \citenamefont {Cruz}, \citenamefont {Albertazzi}, \citenamefont {Bamford},
  \citenamefont {Bell}, \citenamefont {Cross}, \citenamefont {Fraschetti},
  \citenamefont {Graham}, \citenamefont {Hara}, \citenamefont {Kozlowski} \emph
  {et~al.}}]{rigby2018electron}%
  \BibitemOpen
  \bibfield  {author} {\bibinfo {author} {\bibfnamefont {A.}~\bibnamefont
  {Rigby}}, \bibinfo {author} {\bibfnamefont {F.}~\bibnamefont {Cruz}},
  \bibinfo {author} {\bibfnamefont {B.}~\bibnamefont {Albertazzi}}, \bibinfo
  {author} {\bibfnamefont {R.}~\bibnamefont {Bamford}}, \bibinfo {author}
  {\bibfnamefont {A.~R.}\ \bibnamefont {Bell}}, \bibinfo {author}
  {\bibfnamefont {J.~E.}\ \bibnamefont {Cross}}, \bibinfo {author}
  {\bibfnamefont {F.}~\bibnamefont {Fraschetti}}, \bibinfo {author}
  {\bibfnamefont {P.}~\bibnamefont {Graham}}, \bibinfo {author} {\bibfnamefont
  {Y.}~\bibnamefont {Hara}}, \bibinfo {author} {\bibfnamefont {P.~M.}\
  \bibnamefont {Kozlowski}}, \emph {et~al.},\ }\href@noop {} {\bibfield
  {journal} {\bibinfo  {journal} {Nature Physics}\ }\textbf {\bibinfo {volume}
  {14}},\ \bibinfo {pages} {475} (\bibinfo {year} {2018})}\BibitemShut
  {NoStop}%
\bibitem [{\citenamefont {Ahmed}\ \emph {et~al.}(2013)\citenamefont {Ahmed},
  \citenamefont {Dieckmann}, \citenamefont {Romagnani}, \citenamefont {Doria},
  \citenamefont {Sarri}, \citenamefont {Cerchez}, \citenamefont {Ianni},
  \citenamefont {Kourakis}, \citenamefont {Giesecke}, \citenamefont {Notley}
  \emph {et~al.}}]{ahmed2013time}%
  \BibitemOpen
  \bibfield  {author} {\bibinfo {author} {\bibfnamefont {H.}~\bibnamefont
  {Ahmed}}, \bibinfo {author} {\bibfnamefont {M.~E.}\ \bibnamefont
  {Dieckmann}}, \bibinfo {author} {\bibfnamefont {L.}~\bibnamefont
  {Romagnani}}, \bibinfo {author} {\bibfnamefont {D.}~\bibnamefont {Doria}},
  \bibinfo {author} {\bibfnamefont {G.}~\bibnamefont {Sarri}}, \bibinfo
  {author} {\bibfnamefont {M.}~\bibnamefont {Cerchez}}, \bibinfo {author}
  {\bibfnamefont {E.}~\bibnamefont {Ianni}}, \bibinfo {author} {\bibfnamefont
  {I.}~\bibnamefont {Kourakis}}, \bibinfo {author} {\bibfnamefont {A.~L.}\
  \bibnamefont {Giesecke}}, \bibinfo {author} {\bibfnamefont {M.}~\bibnamefont
  {Notley}}, \emph {et~al.},\ }\href@noop {} {\bibfield  {journal} {\bibinfo
  {journal} {Physical Review Letters}\ }\textbf {\bibinfo {volume} {110}},\
  \bibinfo {pages} {205001} (\bibinfo {year} {2013})}\BibitemShut {NoStop}%
\bibitem [{\citenamefont {Jiao}\ \emph {et~al.}(2019)\citenamefont {Jiao},
  \citenamefont {He}, \citenamefont {Zhuo}, \citenamefont {Qiao}, \citenamefont
  {Yu}, \citenamefont {Zhang}, \citenamefont {Deng}, \citenamefont {Lu},
  \citenamefont {Zhou}, \citenamefont {Wang} \emph
  {et~al.}}]{jiao2019experimental}%
  \BibitemOpen
  \bibfield  {author} {\bibinfo {author} {\bibfnamefont {J.}~\bibnamefont
  {Jiao}}, \bibinfo {author} {\bibfnamefont {S.}~\bibnamefont {He}}, \bibinfo
  {author} {\bibfnamefont {H.}~\bibnamefont {Zhuo}}, \bibinfo {author}
  {\bibfnamefont {B.}~\bibnamefont {Qiao}}, \bibinfo {author} {\bibfnamefont
  {M.}~\bibnamefont {Yu}}, \bibinfo {author} {\bibfnamefont {B.}~\bibnamefont
  {Zhang}}, \bibinfo {author} {\bibfnamefont {Z.}~\bibnamefont {Deng}},
  \bibinfo {author} {\bibfnamefont {F.}~\bibnamefont {Lu}}, \bibinfo {author}
  {\bibfnamefont {K.}~\bibnamefont {Zhou}}, \bibinfo {author} {\bibfnamefont
  {X.}~\bibnamefont {Wang}}, \emph {et~al.},\ }\href@noop {} {\bibfield
  {journal} {\bibinfo  {journal} {The Astrophysical Journal Letters}\ }\textbf
  {\bibinfo {volume} {883}},\ \bibinfo {pages} {L37} (\bibinfo {year}
  {2019})}\BibitemShut {NoStop}%
\bibitem [{\citenamefont {Albertazzi}\ \emph {et~al.}(2013)\citenamefont
  {Albertazzi}, \citenamefont {B{\'e}ard}, \citenamefont {Ciardi},
  \citenamefont {Vinci}, \citenamefont {Albrecht}, \citenamefont {Billette},
  \citenamefont {Burris-Mog}, \citenamefont {Chen}, \citenamefont {Da~Silva},
  \citenamefont {Dittrich} \emph {et~al.}}]{albertazzi2013production}%
  \BibitemOpen
  \bibfield  {author} {\bibinfo {author} {\bibfnamefont {B.}~\bibnamefont
  {Albertazzi}}, \bibinfo {author} {\bibfnamefont {J.}~\bibnamefont
  {B{\'e}ard}}, \bibinfo {author} {\bibfnamefont {A.}~\bibnamefont {Ciardi}},
  \bibinfo {author} {\bibfnamefont {T.}~\bibnamefont {Vinci}}, \bibinfo
  {author} {\bibfnamefont {J.}~\bibnamefont {Albrecht}}, \bibinfo {author}
  {\bibfnamefont {J.}~\bibnamefont {Billette}}, \bibinfo {author}
  {\bibfnamefont {T.}~\bibnamefont {Burris-Mog}}, \bibinfo {author}
  {\bibfnamefont {S.}~\bibnamefont {Chen}}, \bibinfo {author} {\bibfnamefont
  {D.}~\bibnamefont {Da~Silva}}, \bibinfo {author} {\bibfnamefont
  {S.}~\bibnamefont {Dittrich}}, \emph {et~al.},\ }\href@noop {} {\bibfield
  {journal} {\bibinfo  {journal} {Review of Scientific Instruments}\ }\textbf
  {\bibinfo {volume} {84}},\ \bibinfo {pages} {043505} (\bibinfo {year}
  {2013})}\BibitemShut {NoStop}%
\bibitem [{\citenamefont {Yao}\ \emph {et~al.}(2019)\citenamefont {Yao},
  \citenamefont {Qiao}, \citenamefont {Zhao}, \citenamefont {Lei},
  \citenamefont {Zhang}, \citenamefont {Zhou}, \citenamefont {Zhu},\ and\
  \citenamefont {He}}]{yao2019kinetic}%
  \BibitemOpen
  \bibfield  {author} {\bibinfo {author} {\bibfnamefont {W.}~\bibnamefont
  {Yao}}, \bibinfo {author} {\bibfnamefont {B.}~\bibnamefont {Qiao}}, \bibinfo
  {author} {\bibfnamefont {Z.}~\bibnamefont {Zhao}}, \bibinfo {author}
  {\bibfnamefont {Z.}~\bibnamefont {Lei}}, \bibinfo {author} {\bibfnamefont
  {H.}~\bibnamefont {Zhang}}, \bibinfo {author} {\bibfnamefont
  {C.}~\bibnamefont {Zhou}}, \bibinfo {author} {\bibfnamefont {S.}~\bibnamefont
  {Zhu}},\ and\ \bibinfo {author} {\bibfnamefont {X.}~\bibnamefont {He}},\
  }\href@noop {} {\bibfield  {journal} {\bibinfo  {journal} {The Astrophysical
  Journal}\ }\textbf {\bibinfo {volume} {876}},\ \bibinfo {pages} {2} (\bibinfo
  {year} {2019})}\BibitemShut {NoStop}%
\bibitem [{\citenamefont {Schaeffer}\ \emph {et~al.}(2020)\citenamefont
  {Schaeffer}, \citenamefont {Fox}, \citenamefont {Matteucci}, \citenamefont
  {Lezhnin}, \citenamefont {Bhattacharjee},\ and\ \citenamefont
  {Germaschewski}}]{schaeffer2020kinetic}%
  \BibitemOpen
  \bibfield  {author} {\bibinfo {author} {\bibfnamefont {D.}~\bibnamefont
  {Schaeffer}}, \bibinfo {author} {\bibfnamefont {W.}~\bibnamefont {Fox}},
  \bibinfo {author} {\bibfnamefont {J.}~\bibnamefont {Matteucci}}, \bibinfo
  {author} {\bibfnamefont {K.}~\bibnamefont {Lezhnin}}, \bibinfo {author}
  {\bibfnamefont {A.}~\bibnamefont {Bhattacharjee}},\ and\ \bibinfo {author}
  {\bibfnamefont {K.}~\bibnamefont {Germaschewski}},\ }\href@noop {} {\bibfield
   {journal} {\bibinfo  {journal} {Physics of Plasmas}\ }\textbf {\bibinfo
  {volume} {27}},\ \bibinfo {pages} {042901} (\bibinfo {year}
  {2020})}\BibitemShut {NoStop}%
\bibitem [{\citenamefont {Higginson}\ \emph {et~al.}(2019)\citenamefont
  {Higginson}, \citenamefont {Korneev}, \citenamefont {Ruyer}, \citenamefont
  {Riquier}, \citenamefont {Moreno}, \citenamefont {B{\'e}ard}, \citenamefont
  {Chen}, \citenamefont {Grassi}, \citenamefont {Grech}, \citenamefont
  {Gremillet} \emph {et~al.}}]{higginson2019laboratory}%
  \BibitemOpen
  \bibfield  {author} {\bibinfo {author} {\bibfnamefont {D.}~\bibnamefont
  {Higginson}}, \bibinfo {author} {\bibfnamefont {P.}~\bibnamefont {Korneev}},
  \bibinfo {author} {\bibfnamefont {C.}~\bibnamefont {Ruyer}}, \bibinfo
  {author} {\bibfnamefont {R.}~\bibnamefont {Riquier}}, \bibinfo {author}
  {\bibfnamefont {Q.}~\bibnamefont {Moreno}}, \bibinfo {author} {\bibfnamefont
  {J.}~\bibnamefont {B{\'e}ard}}, \bibinfo {author} {\bibfnamefont
  {S.}~\bibnamefont {Chen}}, \bibinfo {author} {\bibfnamefont {A.}~\bibnamefont
  {Grassi}}, \bibinfo {author} {\bibfnamefont {M.}~\bibnamefont {Grech}},
  \bibinfo {author} {\bibfnamefont {L.}~\bibnamefont {Gremillet}}, \emph
  {et~al.},\ }\href@noop {} {\bibfield  {journal} {\bibinfo  {journal}
  {Communications Physics}\ }\textbf {\bibinfo {volume} {2}},\ \bibinfo {pages}
  {1} (\bibinfo {year} {2019})}\BibitemShut {NoStop}%
\bibitem [{\citenamefont {Higginson}\ \emph {et~al.}(2017)\citenamefont
  {Higginson}, \citenamefont {Revet}, \citenamefont {Khiar}, \citenamefont
  {B{\'e}ard}, \citenamefont {Blecher}, \citenamefont {Borghesi}, \citenamefont
  {Burdonov}, \citenamefont {Chen}, \citenamefont {Filippov}, \citenamefont
  {Khaghani} \emph {et~al.}}]{higginson2017detailed}%
  \BibitemOpen
  \bibfield  {author} {\bibinfo {author} {\bibfnamefont {D.}~\bibnamefont
  {Higginson}}, \bibinfo {author} {\bibfnamefont {G.}~\bibnamefont {Revet}},
  \bibinfo {author} {\bibfnamefont {B.}~\bibnamefont {Khiar}}, \bibinfo
  {author} {\bibfnamefont {J.}~\bibnamefont {B{\'e}ard}}, \bibinfo {author}
  {\bibfnamefont {M.}~\bibnamefont {Blecher}}, \bibinfo {author} {\bibfnamefont
  {M.}~\bibnamefont {Borghesi}}, \bibinfo {author} {\bibfnamefont
  {K.}~\bibnamefont {Burdonov}}, \bibinfo {author} {\bibfnamefont
  {S.}~\bibnamefont {Chen}}, \bibinfo {author} {\bibfnamefont {E.}~\bibnamefont
  {Filippov}}, \bibinfo {author} {\bibfnamefont {D.}~\bibnamefont {Khaghani}},
  \emph {et~al.},\ }\href@noop {} {\bibfield  {journal} {\bibinfo  {journal}
  {High Energy Density Physics}\ }\textbf {\bibinfo {volume} {23}},\ \bibinfo
  {pages} {48} (\bibinfo {year} {2017})}\BibitemShut {NoStop}%
\bibitem [{\citenamefont {Khiar}\ \emph {et~al.}(2019)\citenamefont {Khiar},
  \citenamefont {Revet}, \citenamefont {Ciardi}, \citenamefont {Burdonov},
  \citenamefont {Filippov}, \citenamefont {B{\'e}ard}, \citenamefont {Cerchez},
  \citenamefont {Chen}, \citenamefont {Gangolf}, \citenamefont {Makarov} \emph
  {et~al.}}]{khiar2019laser}%
  \BibitemOpen
  \bibfield  {author} {\bibinfo {author} {\bibfnamefont {B.}~\bibnamefont
  {Khiar}}, \bibinfo {author} {\bibfnamefont {G.}~\bibnamefont {Revet}},
  \bibinfo {author} {\bibfnamefont {A.}~\bibnamefont {Ciardi}}, \bibinfo
  {author} {\bibfnamefont {K.}~\bibnamefont {Burdonov}}, \bibinfo {author}
  {\bibfnamefont {E.}~\bibnamefont {Filippov}}, \bibinfo {author}
  {\bibfnamefont {J.}~\bibnamefont {B{\'e}ard}}, \bibinfo {author}
  {\bibfnamefont {M.}~\bibnamefont {Cerchez}}, \bibinfo {author} {\bibfnamefont
  {S.}~\bibnamefont {Chen}}, \bibinfo {author} {\bibfnamefont {T.}~\bibnamefont
  {Gangolf}}, \bibinfo {author} {\bibfnamefont {S.}~\bibnamefont {Makarov}},
  \emph {et~al.},\ }\href@noop {} {\bibfield  {journal} {\bibinfo  {journal}
  {Physical Review Letters}\ }\textbf {\bibinfo {volume} {123}},\ \bibinfo
  {pages} {205001} (\bibinfo {year} {2019})}\BibitemShut {NoStop}%
\bibitem [{\citenamefont {Filippov}\ \emph {et~al.}(2020)\citenamefont
  {Filippov}, \citenamefont {Makarov}, \citenamefont {Burdonov}, \citenamefont
  {Yao}, \citenamefont {Revet}, \citenamefont {B{\'e}ard}, \citenamefont
  {Bola{\~n}os}, \citenamefont {Chen}, \citenamefont {Guediche}, \citenamefont
  {Hare} \emph {et~al.}}]{filippov2020enhanced}%
  \BibitemOpen
  \bibfield  {author} {\bibinfo {author} {\bibfnamefont {E.}~\bibnamefont
  {Filippov}}, \bibinfo {author} {\bibfnamefont {S.}~\bibnamefont {Makarov}},
  \bibinfo {author} {\bibfnamefont {K.}~\bibnamefont {Burdonov}}, \bibinfo
  {author} {\bibfnamefont {W.}~\bibnamefont {Yao}}, \bibinfo {author}
  {\bibfnamefont {G.}~\bibnamefont {Revet}}, \bibinfo {author} {\bibfnamefont
  {J.}~\bibnamefont {B{\'e}ard}}, \bibinfo {author} {\bibfnamefont
  {S.}~\bibnamefont {Bola{\~n}os}}, \bibinfo {author} {\bibfnamefont
  {S.}~\bibnamefont {Chen}}, \bibinfo {author} {\bibfnamefont {A.}~\bibnamefont
  {Guediche}}, \bibinfo {author} {\bibfnamefont {J.}~\bibnamefont {Hare}},
  \emph {et~al.},\ }\href@noop {} {\bibfield  {journal} {\bibinfo  {journal}
  {arXiv preprint arXiv:2006.12424}\ } (\bibinfo {year} {2020})}\BibitemShut
  {NoStop}%
\bibitem [{\citenamefont {Giagkiozis}\ \emph {et~al.}(2017)\citenamefont
  {Giagkiozis}, \citenamefont {Walker}, \citenamefont {Pope},\ and\
  \citenamefont {Collinson}}]{giagkiozis2017validation}%
  \BibitemOpen
  \bibfield  {author} {\bibinfo {author} {\bibfnamefont {S.}~\bibnamefont
  {Giagkiozis}}, \bibinfo {author} {\bibfnamefont {S.~N.}\ \bibnamefont
  {Walker}}, \bibinfo {author} {\bibfnamefont {S.~A.}\ \bibnamefont {Pope}},\
  and\ \bibinfo {author} {\bibfnamefont {G.}~\bibnamefont {Collinson}},\
  }\href@noop {} {\bibfield  {journal} {\bibinfo  {journal} {Journal of
  Geophysical Research: Space Physics}\ }\textbf {\bibinfo {volume} {122}},\
  \bibinfo {pages} {8632} (\bibinfo {year} {2017})}\BibitemShut {NoStop}%
\bibitem [{\citenamefont {Faenov}\ \emph {et~al.}(1994)\citenamefont {Faenov},
  \citenamefont {Pikuz}, \citenamefont {Erko}, \citenamefont {Bryunetkin} \emph
  {et~al.}}]{Faenov:1994}%
  \BibitemOpen
  \bibfield  {author} {\bibinfo {author} {\bibfnamefont {A.~Y.}\ \bibnamefont
  {Faenov}}, \bibinfo {author} {\bibfnamefont {S.~A.}\ \bibnamefont {Pikuz}},
  \bibinfo {author} {\bibfnamefont {A.~I.}\ \bibnamefont {Erko}}, \bibinfo
  {author} {\bibfnamefont {B.~A.}\ \bibnamefont {Bryunetkin}}, \emph {et~al.},\
  }\href@noop {} {\bibfield  {journal} {\bibinfo  {journal} {Phys. Scr.}\
  }\textbf {\bibinfo {volume} {50}},\ \bibinfo {pages} {333} (\bibinfo {year}
  {1994})}\BibitemShut {NoStop}%
\bibitem [{\citenamefont {Ryazantsev}\ \emph {et~al.}(2015)\citenamefont
  {Ryazantsev}, \citenamefont {Skobelev}, \citenamefont {Faenov}, \citenamefont
  {Pikuz}, \citenamefont {Grum-Grzhimailo},\ and\ \citenamefont
  {Pikuz}}]{Ryazantsev2015}%
  \BibitemOpen
  \bibfield  {author} {\bibinfo {author} {\bibfnamefont {S.~N.}\ \bibnamefont
  {Ryazantsev}}, \bibinfo {author} {\bibfnamefont {I.~Y.}\ \bibnamefont
  {Skobelev}}, \bibinfo {author} {\bibfnamefont {A.~Y.}\ \bibnamefont
  {Faenov}}, \bibinfo {author} {\bibfnamefont {T.~A.}\ \bibnamefont {Pikuz}},
  \bibinfo {author} {\bibfnamefont {A.~N.}\ \bibnamefont {Grum-Grzhimailo}},\
  and\ \bibinfo {author} {\bibfnamefont {S.~A.}\ \bibnamefont {Pikuz}},\ }\href
  {https://doi.org/10.1134/S0021364015230149} {\bibfield  {journal} {\bibinfo
  {journal} {JETP Letters}\ }\textbf {\bibinfo {volume} {102}},\ \bibinfo
  {pages} {707} (\bibinfo {year} {2015})}\BibitemShut {NoStop}%
\bibitem [{\citenamefont {Filippov}\ \emph {et~al.}(2019)\citenamefont
  {Filippov}, \citenamefont {Skobelev}, \citenamefont {Revet}, \citenamefont
  {Chen}, \citenamefont {Khiar}, \citenamefont {Ciardi}, \citenamefont
  {Khaghani}, \citenamefont {Higginson}, \citenamefont {Pikuz},\ and\
  \citenamefont {Fuchs}}]{Filippov2019}%
  \BibitemOpen
  \bibfield  {author} {\bibinfo {author} {\bibfnamefont {E.~D.}\ \bibnamefont
  {Filippov}}, \bibinfo {author} {\bibfnamefont {I.~Y.}\ \bibnamefont
  {Skobelev}}, \bibinfo {author} {\bibfnamefont {G.}~\bibnamefont {Revet}},
  \bibinfo {author} {\bibfnamefont {S.~N.}\ \bibnamefont {Chen}}, \bibinfo
  {author} {\bibfnamefont {B.}~\bibnamefont {Khiar}}, \bibinfo {author}
  {\bibfnamefont {A.}~\bibnamefont {Ciardi}}, \bibinfo {author} {\bibfnamefont
  {D.}~\bibnamefont {Khaghani}}, \bibinfo {author} {\bibfnamefont {D.~P.}\
  \bibnamefont {Higginson}}, \bibinfo {author} {\bibfnamefont {S.~A.}\
  \bibnamefont {Pikuz}},\ and\ \bibinfo {author} {\bibfnamefont
  {J.}~\bibnamefont {Fuchs}},\ }\href {https://doi.org/10.1063/1.5124350}
  {\bibfield  {journal} {\bibinfo  {journal} {Matter and Radiation at
  Extremes}\ }\textbf {\bibinfo {volume} {4}},\ \bibinfo {pages} {064402}
  (\bibinfo {year} {2019})}\BibitemShut {NoStop}%
\bibitem [{\citenamefont {Wilks}\ \emph {et~al.}(1992)\citenamefont {Wilks},
  \citenamefont {Kruer}, \citenamefont {Tabak},\ and\ \citenamefont
  {Langdon}}]{wilks1992absorption}%
  \BibitemOpen
  \bibfield  {author} {\bibinfo {author} {\bibfnamefont {S.}~\bibnamefont
  {Wilks}}, \bibinfo {author} {\bibfnamefont {W.}~\bibnamefont {Kruer}},
  \bibinfo {author} {\bibfnamefont {M.}~\bibnamefont {Tabak}},\ and\ \bibinfo
  {author} {\bibfnamefont {A.}~\bibnamefont {Langdon}},\ }\href@noop {}
  {\bibfield  {journal} {\bibinfo  {journal} {Physical review letters}\
  }\textbf {\bibinfo {volume} {69}},\ \bibinfo {pages} {1383} (\bibinfo {year}
  {1992})}\BibitemShut {NoStop}%
\bibitem [{\citenamefont {Froula}\ \emph {et~al.}(2011)\citenamefont {Froula},
  \citenamefont {Luhmann~Jr}, \citenamefont {Sheffield},\ and\ \citenamefont
  {Glenzer}}]{froula2011plasma}%
  \BibitemOpen
  \bibfield  {author} {\bibinfo {author} {\bibfnamefont {D.~H.}\ \bibnamefont
  {Froula}}, \bibinfo {author} {\bibfnamefont {N.~C.}\ \bibnamefont
  {Luhmann~Jr}}, \bibinfo {author} {\bibfnamefont {J.}~\bibnamefont
  {Sheffield}},\ and\ \bibinfo {author} {\bibfnamefont {S.~H.}\ \bibnamefont
  {Glenzer}},\ }\href@noop {} {\emph {\bibinfo {title} {Plasma scattering of
  electromagnetic radiation: theory and measurement techniques}}}\ (\bibinfo
  {publisher} {Elsevier},\ \bibinfo {year} {2011})\BibitemShut {NoStop}%
\bibitem [{\citenamefont {Froula}\ \emph {et~al.}(2007)\citenamefont {Froula},
  \citenamefont {Ross}, \citenamefont {Pollock}, \citenamefont {Davis},
  \citenamefont {James}, \citenamefont {Divol}, \citenamefont {Edwards},
  \citenamefont {Offenberger}, \citenamefont {Price}, \citenamefont {Town}
  \emph {et~al.}}]{froula2007quenching}%
  \BibitemOpen
  \bibfield  {author} {\bibinfo {author} {\bibfnamefont {D.}~\bibnamefont
  {Froula}}, \bibinfo {author} {\bibfnamefont {J.}~\bibnamefont {Ross}},
  \bibinfo {author} {\bibfnamefont {B.}~\bibnamefont {Pollock}}, \bibinfo
  {author} {\bibfnamefont {P.}~\bibnamefont {Davis}}, \bibinfo {author}
  {\bibfnamefont {A.}~\bibnamefont {James}}, \bibinfo {author} {\bibfnamefont
  {L.}~\bibnamefont {Divol}}, \bibinfo {author} {\bibfnamefont
  {M.}~\bibnamefont {Edwards}}, \bibinfo {author} {\bibfnamefont
  {A.}~\bibnamefont {Offenberger}}, \bibinfo {author} {\bibfnamefont
  {D.}~\bibnamefont {Price}}, \bibinfo {author} {\bibfnamefont
  {R.}~\bibnamefont {Town}}, \emph {et~al.},\ }\href@noop {} {\bibfield
  {journal} {\bibinfo  {journal} {Physical Review Letters}\ }\textbf {\bibinfo
  {volume} {98}},\ \bibinfo {pages} {135001} (\bibinfo {year}
  {2007})}\BibitemShut {NoStop}%
\bibitem [{\citenamefont {Man{\v{c}}i{\'c}}\ \emph {et~al.}(2008)\citenamefont
  {Man{\v{c}}i{\'c}}, \citenamefont {Fuchs}, \citenamefont {Antici},
  \citenamefont {Gaillard},\ and\ \citenamefont
  {Audebert}}]{manvcic2008absolute}%
  \BibitemOpen
  \bibfield  {author} {\bibinfo {author} {\bibfnamefont {A.}~\bibnamefont
  {Man{\v{c}}i{\'c}}}, \bibinfo {author} {\bibfnamefont {J.}~\bibnamefont
  {Fuchs}}, \bibinfo {author} {\bibfnamefont {P.}~\bibnamefont {Antici}},
  \bibinfo {author} {\bibfnamefont {S.}~\bibnamefont {Gaillard}},\ and\
  \bibinfo {author} {\bibfnamefont {P.}~\bibnamefont {Audebert}},\ }\href@noop
  {} {\bibfield  {journal} {\bibinfo  {journal} {Review of Scientific
  Instruments}\ }\textbf {\bibinfo {volume} {79}},\ \bibinfo {pages} {073301}
  (\bibinfo {year} {2008})}\BibitemShut {NoStop}%
\bibitem [{\citenamefont {Fryxell}\ \emph {et~al.}(2000)\citenamefont
  {Fryxell}, \citenamefont {Olson}, \citenamefont {Ricker}, \citenamefont
  {Timmes}, \citenamefont {Zingale}, \citenamefont {Lamb}, \citenamefont
  {MacNeice}, \citenamefont {Rosner}, \citenamefont {Truran},\ and\
  \citenamefont {Tufo}}]{fryxell2000flash}%
  \BibitemOpen
  \bibfield  {author} {\bibinfo {author} {\bibfnamefont {B.}~\bibnamefont
  {Fryxell}}, \bibinfo {author} {\bibfnamefont {K.}~\bibnamefont {Olson}},
  \bibinfo {author} {\bibfnamefont {P.}~\bibnamefont {Ricker}}, \bibinfo
  {author} {\bibfnamefont {F.}~\bibnamefont {Timmes}}, \bibinfo {author}
  {\bibfnamefont {M.}~\bibnamefont {Zingale}}, \bibinfo {author} {\bibfnamefont
  {D.}~\bibnamefont {Lamb}}, \bibinfo {author} {\bibfnamefont {P.}~\bibnamefont
  {MacNeice}}, \bibinfo {author} {\bibfnamefont {R.}~\bibnamefont {Rosner}},
  \bibinfo {author} {\bibfnamefont {J.}~\bibnamefont {Truran}},\ and\ \bibinfo
  {author} {\bibfnamefont {H.}~\bibnamefont {Tufo}},\ }\href@noop {} {\bibfield
   {journal} {\bibinfo  {journal} {The Astrophysical Journal Supplement
  Series}\ }\textbf {\bibinfo {volume} {131}},\ \bibinfo {pages} {273}
  (\bibinfo {year} {2000})}\BibitemShut {NoStop}%
\bibitem [{\citenamefont {Kemp}\ and\ \citenamefont {Meyer-ter
  Vehn}(1998)}]{kemp1998equation}%
  \BibitemOpen
  \bibfield  {author} {\bibinfo {author} {\bibfnamefont {A.}~\bibnamefont
  {Kemp}}\ and\ \bibinfo {author} {\bibfnamefont {J.}~\bibnamefont {Meyer-ter
  Vehn}},\ }\href@noop {} {\bibfield  {journal} {\bibinfo  {journal} {Nuclear
  Instruments and Methods in Physics Research Section A: Accelerators,
  Spectrometers, Detectors and Associated Equipment}\ }\textbf {\bibinfo
  {volume} {415}},\ \bibinfo {pages} {674} (\bibinfo {year}
  {1998})}\BibitemShut {NoStop}%
\bibitem [{\citenamefont {Haines}(1986)}]{haines1986magnetic}%
  \BibitemOpen
  \bibfield  {author} {\bibinfo {author} {\bibfnamefont {M.}~\bibnamefont
  {Haines}},\ }\href@noop {} {\bibfield  {journal} {\bibinfo  {journal}
  {Canadian Journal of Physics}\ }\textbf {\bibinfo {volume} {64}},\ \bibinfo
  {pages} {912} (\bibinfo {year} {1986})}\BibitemShut {NoStop}%
\bibitem [{\citenamefont {Li}\ \emph {et~al.}(2006)\citenamefont {Li},
  \citenamefont {S\'eguin}, \citenamefont {Frenje}, \citenamefont {Rygg},
  \citenamefont {Petrasso}, \citenamefont {Town}, \citenamefont {Amendt},
  \citenamefont {Hatchett}, \citenamefont {Landen}, \citenamefont {Mackinnon},
  \citenamefont {Patel}, \citenamefont {Smalyuk}, \citenamefont {Sangster},\
  and\ \citenamefont {Knauer}}]{li_2006}%
  \BibitemOpen
  \bibfield  {author} {\bibinfo {author} {\bibfnamefont {C.~K.}\ \bibnamefont
  {Li}}, \bibinfo {author} {\bibfnamefont {F.~H.}\ \bibnamefont {S\'eguin}},
  \bibinfo {author} {\bibfnamefont {J.~A.}\ \bibnamefont {Frenje}}, \bibinfo
  {author} {\bibfnamefont {J.~R.}\ \bibnamefont {Rygg}}, \bibinfo {author}
  {\bibfnamefont {R.~D.}\ \bibnamefont {Petrasso}}, \bibinfo {author}
  {\bibfnamefont {R.~P.~J.}\ \bibnamefont {Town}}, \bibinfo {author}
  {\bibfnamefont {P.~A.}\ \bibnamefont {Amendt}}, \bibinfo {author}
  {\bibfnamefont {S.~P.}\ \bibnamefont {Hatchett}}, \bibinfo {author}
  {\bibfnamefont {O.~L.}\ \bibnamefont {Landen}}, \bibinfo {author}
  {\bibfnamefont {A.~J.}\ \bibnamefont {Mackinnon}}, \bibinfo {author}
  {\bibfnamefont {P.~K.}\ \bibnamefont {Patel}}, \bibinfo {author}
  {\bibfnamefont {V.~A.}\ \bibnamefont {Smalyuk}}, \bibinfo {author}
  {\bibfnamefont {T.~C.}\ \bibnamefont {Sangster}},\ and\ \bibinfo {author}
  {\bibfnamefont {J.~P.}\ \bibnamefont {Knauer}},\ }\href
  {https://doi.org/10.1103/PhysRevLett.97.135003} {\bibfield  {journal}
  {\bibinfo  {journal} {Phys. Rev. Lett.}\ }\textbf {\bibinfo {volume} {97}},\
  \bibinfo {pages} {135003} (\bibinfo {year} {2006})}\BibitemShut {NoStop}%
\bibitem [{\citenamefont {Gao}\ \emph {et~al.}(2015)\citenamefont {Gao},
  \citenamefont {Nilson}, \citenamefont {Igumenshchev}, \citenamefont {Haines},
  \citenamefont {Froula}, \citenamefont {Betti},\ and\ \citenamefont
  {Meyerhofer}}]{Gao2015}%
  \BibitemOpen
  \bibfield  {author} {\bibinfo {author} {\bibfnamefont {L.}~\bibnamefont
  {Gao}}, \bibinfo {author} {\bibfnamefont {P.~M.}\ \bibnamefont {Nilson}},
  \bibinfo {author} {\bibfnamefont {I.~V.}\ \bibnamefont {Igumenshchev}},
  \bibinfo {author} {\bibfnamefont {M.~G.}\ \bibnamefont {Haines}}, \bibinfo
  {author} {\bibfnamefont {D.~H.}\ \bibnamefont {Froula}}, \bibinfo {author}
  {\bibfnamefont {R.}~\bibnamefont {Betti}},\ and\ \bibinfo {author}
  {\bibfnamefont {D.~D.}\ \bibnamefont {Meyerhofer}},\ }\href
  {https://doi.org/10.1103/PhysRevLett.114.215003} {\bibfield  {journal}
  {\bibinfo  {journal} {Physical Review Letters}\ }\textbf {\bibinfo {volume}
  {114}},\ \bibinfo {pages} {215003} (\bibinfo {year} {2015})}\BibitemShut
  {NoStop}%
\bibitem [{\citenamefont {Cecchetti}\ \emph {et~al.}(2009)\citenamefont
  {Cecchetti}, \citenamefont {Borghesi}, \citenamefont {Fuchs}, \citenamefont
  {Schurtz}, \citenamefont {Kar}, \citenamefont {Macchi}, \citenamefont
  {Romagnani}, \citenamefont {Wilson}, \citenamefont {Antici}, \citenamefont
  {Jung}, \citenamefont {Osterholtz}, \citenamefont {Pipahl}, \citenamefont
  {Willi}, \citenamefont {Schiavi}, \citenamefont {Notley},\ and\ \citenamefont
  {Neely}}]{cecchetti_2009}%
  \BibitemOpen
  \bibfield  {author} {\bibinfo {author} {\bibfnamefont {C.~A.}\ \bibnamefont
  {Cecchetti}}, \bibinfo {author} {\bibfnamefont {M.}~\bibnamefont {Borghesi}},
  \bibinfo {author} {\bibfnamefont {J.}~\bibnamefont {Fuchs}}, \bibinfo
  {author} {\bibfnamefont {G.}~\bibnamefont {Schurtz}}, \bibinfo {author}
  {\bibfnamefont {S.}~\bibnamefont {Kar}}, \bibinfo {author} {\bibfnamefont
  {A.}~\bibnamefont {Macchi}}, \bibinfo {author} {\bibfnamefont
  {L.}~\bibnamefont {Romagnani}}, \bibinfo {author} {\bibfnamefont {P.~A.}\
  \bibnamefont {Wilson}}, \bibinfo {author} {\bibfnamefont {P.}~\bibnamefont
  {Antici}}, \bibinfo {author} {\bibfnamefont {R.}~\bibnamefont {Jung}},
  \bibinfo {author} {\bibfnamefont {J.}~\bibnamefont {Osterholtz}}, \bibinfo
  {author} {\bibfnamefont {C.~A.}\ \bibnamefont {Pipahl}}, \bibinfo {author}
  {\bibfnamefont {O.}~\bibnamefont {Willi}}, \bibinfo {author} {\bibfnamefont
  {A.}~\bibnamefont {Schiavi}}, \bibinfo {author} {\bibfnamefont
  {M.}~\bibnamefont {Notley}},\ and\ \bibinfo {author} {\bibfnamefont
  {D.}~\bibnamefont {Neely}},\ }\href {https://doi.org/10.1063/1.3097899}
  {\bibfield  {journal} {\bibinfo  {journal} {Physics of Plasmas}\ }\textbf
  {\bibinfo {volume} {16}},\ \bibinfo {pages} {043102} (\bibinfo {year}
  {2009})}\BibitemShut {NoStop}%
\bibitem [{\citenamefont {Derouillat}\ \emph {et~al.}(2018)\citenamefont
  {Derouillat}, \citenamefont {Beck}, \citenamefont {P{\'e}rez}, \citenamefont
  {Vinci}, \citenamefont {Chiaramello}, \citenamefont {Grassi}, \citenamefont
  {Fl{\'e}}, \citenamefont {Bouchard}, \citenamefont {Plotnikov}, \citenamefont
  {Aunai} \emph {et~al.}}]{derouillat2018smilei}%
  \BibitemOpen
  \bibfield  {author} {\bibinfo {author} {\bibfnamefont {J.}~\bibnamefont
  {Derouillat}}, \bibinfo {author} {\bibfnamefont {A.}~\bibnamefont {Beck}},
  \bibinfo {author} {\bibfnamefont {F.}~\bibnamefont {P{\'e}rez}}, \bibinfo
  {author} {\bibfnamefont {T.}~\bibnamefont {Vinci}}, \bibinfo {author}
  {\bibfnamefont {M.}~\bibnamefont {Chiaramello}}, \bibinfo {author}
  {\bibfnamefont {A.}~\bibnamefont {Grassi}}, \bibinfo {author} {\bibfnamefont
  {M.}~\bibnamefont {Fl{\'e}}}, \bibinfo {author} {\bibfnamefont
  {G.}~\bibnamefont {Bouchard}}, \bibinfo {author} {\bibfnamefont
  {I.}~\bibnamefont {Plotnikov}}, \bibinfo {author} {\bibfnamefont
  {N.}~\bibnamefont {Aunai}}, \emph {et~al.},\ }\href@noop {} {\bibfield
  {journal} {\bibinfo  {journal} {Computer Physics Communications}\ }\textbf
  {\bibinfo {volume} {222}},\ \bibinfo {pages} {351} (\bibinfo {year}
  {2018})}\BibitemShut {NoStop}%
\bibitem [{\citenamefont {Matsukiyo}\ and\ \citenamefont
  {Scholer}(2003)}]{matsukiyo2003modified}%
  \BibitemOpen
  \bibfield  {author} {\bibinfo {author} {\bibfnamefont {S.}~\bibnamefont
  {Matsukiyo}}\ and\ \bibinfo {author} {\bibfnamefont {M.}~\bibnamefont
  {Scholer}},\ }\href@noop {} {\bibfield  {journal} {\bibinfo  {journal}
  {Journal of Geophysical Research: Space Physics}\ }\textbf {\bibinfo {volume}
  {108}} (\bibinfo {year} {2003})}\BibitemShut {NoStop}%
\bibitem [{\citenamefont {Woods}(1971)}]{woods1972shock}%
  \BibitemOpen
  \bibfield  {author} {\bibinfo {author} {\bibfnamefont {L.}~\bibnamefont
  {Woods}},\ }\href@noop {} {\emph {\bibinfo {title} {Shock Waves in
  Collisionless Plasmas}}},\ edited by\ \bibinfo {editor} {\bibfnamefont
  {D.}~\bibnamefont {Tidman}}\ and\ \bibinfo {editor} {\bibfnamefont
  {N.}~\bibnamefont {Krall}}\ (\bibinfo  {publisher} {Wiley, New York},\
  \bibinfo {year} {1971})\BibitemShut {NoStop}%
\bibitem [{\citenamefont {Burgess}\ and\ \citenamefont
  {Scholer}(2007)}]{burgess2007shock}%
  \BibitemOpen
  \bibfield  {author} {\bibinfo {author} {\bibfnamefont {D.}~\bibnamefont
  {Burgess}}\ and\ \bibinfo {author} {\bibfnamefont {M.}~\bibnamefont
  {Scholer}},\ }\href@noop {} {\bibfield  {journal} {\bibinfo  {journal}
  {Physics of Plasmas}\ }\textbf {\bibinfo {volume} {14}},\ \bibinfo {pages}
  {012108} (\bibinfo {year} {2007})}\BibitemShut {NoStop}%
\end{thebibliography}%

\end{document}